\documentclass[12pt,a4paper]{article}
\usepackage{latexsym,amssymb,epsfig}

\newcommand{\bm}[1]{\mbox{\boldmath$#1$}}
\def\0{\phantom{0}}

\begin{document}
\pagenumbering{arabic}
\baselineskip25pt

\begin{center}
{\bf \large Comprehensive study of the vapour-liquid equilibria of the pure two-centre
Lennard-Jones plus pointquadrupole fluid} \\
\bigskip
\renewcommand{\thefootnote}{\fnsymbol{footnote}}
J\"urgen Stoll, Jadran Vrabec, Hans Hasse\footnote[1]{author for correspondence, Tel.: ++49-711/685-6105, Fax: ++49-711/685-7657, Email: hasse@itt.uni-stuttgart.de } \\
\renewcommand{\thefootnote}{\arabic{footnote}}
Institut f\"ur Technische Thermodynamik und Thermische Verfahrenstechnik, \\
Universit\"at Stuttgart, D-70550 Stuttgart, Germany \\
\vspace{0.6cm}
Johann Fischer \\
Institut f\"ur Land-, Umwelt- und Energietechnik, \\
Universit\"at f\"ur Bodenkultur Wien, A-1190 Wien, Austria
\end{center}
\begin{abstract}
\baselineskip25pt
Results of a systematic investigation of the vapour-liquid equilibria (VLE) of 30 individual two-centre Lennard-Jones plus pointquadrupole model fluids (2CLJQ) are reported over a range of reduced quadrupolar momentum $0 \le Q^{*2} \le 4$ and of reduced elongation $0 \le L^* \le 0.8$. Temperatures investigated are from about 55 \% to about 95 \% of the critical temperature of each fluid. Uniformly the $N\!pT$+Test Particle Method based on molecular dynamics simulations is used for the generation of vapour pressures, saturated densities, and saturated enthalpies. Critical temperatures $T^*_{\rm c}$ and densities $\rho^*_{\rm c}$ are obtained from Guggenheim's equations. Empirical correlations for critical data $T^*_{\rm c}$ and $\rho^*_{\rm c}$ as well as for saturated densities $\rho'^*$, $\rho''^*$, and  vapour pressures $p^*_{\sigma}$ are developed as global functions of the model parameters. In most cases they describe the simulation data within their statistical uncertainties. Critical pressures and acentric factors of the 2CLJQ fluid can be calculated from present correlations. The present simulations are a sound basis for adjustments of the model parameters $Q^{*2}$, $L^*$, $\sigma$, and $\epsilon$ to experimental VLE data of real fluids.
\end{abstract}

\section{Introduction}

Knowledge on vapor-liquid equilibria (VLE) is important in many problems in engineering and natural sciences. Among the different ways to model vapor-liquid equilibria, molecular simulation has the highest potential to yield significant improvements compared to existing phenomenological models, especially in terms of predictive power. However, very significant further efforts will be needed until molecular simulation based models and tools will be sufficiently developed so that their advantages can help process engineers in their real world tasks. 

One of the main problems to be overcome is the lack of intermolecular interaction models describing vapor-liquid equilibra of real fluids with acceptable accuracy. (With "real" fluids we do not refer to non-ideality here, but rather to real world fluids like, e.g., 1-propanol or chlorobenzene). As in most applications mixtures are of interest, it is necessary to model the different pure fluids with a family of compatible interaction models, allowing the application of simple combining rules. Presently the probably most attractive and promising family of interaction models are the Lennard-Jones based models, which have successfully been applied to a number of real fluids. We mention only some. Vrabec et al. \cite{vrabec9578} have adjusted the parameters of the one centre Lennard-Jones fluid (1CLJ) to argon and methane and the two centre Lennard-Jones fluid to ethane \cite{vrabec9688}. Van Leeuwen \cite{leeuwen941} presented parameters for the one centre Lennard-Jones plus dipole fluid (1CLJD) for various real substances. Two centre Lennard-Jones plus dipole (2CLJD) parameters for various refrigerants are reported by Kriebel et al. \cite{kriebel9815} and more recently by L\'{\i}sal et al. \cite{lisal9916}. Finally, the two centre Lennard-Jones plus pointquadrupole (2CLJQ) fluid has successfully been used to describe the thermodynamic properties of carbon dioxide (M\"oller et al. \cite{moeller9435}). Without reference to real fluids, Dubey et al. \cite{dubey9421} have studied phase equilibria of two centre Lennard-Jones dipolar plus quadrupolar fluids (2CLJDQ). O'Shea et al. \cite{oshea9723} have reported phase equilibria for one centre Lennard-Jones plus quadrupole fluids (1CLJQ).

The search for an appropriate interaction model for a given fluid is usually a time consuming process. In general the focus lies on {\it one pure fluid}, where the optimization of the potential model is done by a number of simulations with subsequent variation of the model parameters, cf., e.g., van Leeuwen et al. \cite{leeuwen9518} for methanol. In the present work, we follow a new route to develop interaction models that allows fast adjustments of model parameters to experimental data for a given {\it class of pure fluids}. Here we study quadrupolar fluids, of which carbon dioxide undoubted is the most prominent member. The favourable results obtained by M\"oller et al. \cite{moeller9435} for carbon dioxide with their 2CLJQ model give reason to believe that that model will also be successful for other quadrupolar fluids. This is supported by results from a recent study of Calero et al. on refrigerants \cite{calero981}. The idea we follow is to study the thermodynamic properties of the 2CLJQ model fluid systematically and in detail over a wide range of model parameters. Using reduced coordinates, for the symmetric 2CLJQ fluid only two parameters have to be varied: the dimensionless LJ centre-centre distance $L^*$ and the dimensionless pointquadrupole strenght $Q^{*2}$. This allows to cover the entire parameter space of interest with acceptable accuracy by studying 30 individual 2CLJQ fluids with different numbers for $L^*$ and $Q^{*2}$. The simulation results are correlated empirically in order to be able to interpolate between the discrete pairs of $L^*$ and $Q^{*2}$ respectively.

Based on the results from the present study, it is straightforward to adjust the molecular interaction parameters of the 2CLJQ fluid to experimental data of real quadrupolar fluids. Properties like the critical values of temperature and pressure, the azentric factor or the saturated liquid density and vapour pressure are available as functions of the molecular interaction parameters. Hence, the development of the molecular interaction model for a given substance is not more difficult than the adjustment of parameters of phenomenological thermodynamic models. 

The present work covers the basic molecular simulations for 30 individual 2CLJQ fluids and the development of the correlations together with a discussion of those results. The application to real fluids will be presented separately. 

Different methods for simulating VLE are available. Many of them are based on Monte Carlo techniques (MC) such as the Gibbs Ensemble Monte Carlo by Panagiotopoulos \cite{panagiotopoulos8781}, that is broadly used.  Another method is the Gibbs-Duhem Integration, presented by Kofke \cite{kofke9313,kofke9341}, which uses information from integration of the Clausius-Clapeyron equation. Valleau \cite{valleau9119} and Kiyohara et al. \cite{kiyohara9696} presented MC based Thermodynamic Scaling Methods. Mehta et al. \cite{mehta9513} suggested Pseudo-Ensemble Methods that avoid particle insertion. Recently, Histogram-Reweighting Methods have been applied for the determination of VLE by Kiyohara et al. \cite{kiyohara9733,kiyohara9880} and Conrad et al. \cite{conrad9851}. For simulations near the critical points Finite Size Scaling Methods have been applied successfully by Wilding \cite{wilding9560,wilding9758}. M\"oller et al. \cite{moeller9046} have suggested the $N\!pT+$Test Particle Method.  The $N\!pT+$Test Particle Method can be based both on MC simulations \cite{boda9514} and on molecular dynamics simulations (MD) \cite{moeller9435,lotfi9213,kriebel9538}. The different simulation methods vary in accuracy and computing time; we have chosen the $N\!pT$+Test Particle Method due to our favourable experience with that method especially concerning accuracy.

\section{Investigated models}

In the present investigation we consider pure two centre Lennard-Jones plus pointquadrupole fluids (2CLJQ). It is composed of two identical Lennard-Jones sites a distance $L$ apart (2CLJ) plus a pointquadrupole of momentum $Q$ placed in the geometric centre of the molecule. The full potential $u_{\rm 2CLJQ}$ writes as
\begin{eqnarray}
u_{\rm 2CLJQ}(\bm{r}_{ij},\bm{\omega}_i,\bm{\omega}_j,L,Q^2) = u_{\rm 2CLJ}(\bm{r}_{ij},\bm{\omega}_i,\bm{\omega}_j,L)+u_{\rm Q}(\bm{r}_{ij},\bm{\omega}_i,\bm{\omega}_j,Q^2), \nonumber
\end{eqnarray}
wherein
\begin{eqnarray}
u_{\rm 2CLJ}(\bm{r}_{ij},\bm{\omega}_i,\bm{\omega}_j,L,\sigma,\epsilon)=\sum_{a=1}^{2} \sum_{b=1}^{2} 4\epsilon \left[ \left( \frac{\sigma}{r_{ab}} \right)^{12} - \left( \frac{\sigma}{r_{ab}} \right)^6 \right] \nonumber
\end{eqnarray}
and, as given by Gray et al. \cite{gray84}
\begin{eqnarray}
u_{\rm Q}(\bm{r}_{ij},\bm{\omega}_i,\bm{\omega}_j,Q^2)=\frac{3}{4} \frac{Q^2}{\left|\bm{r}_{ij}\right|^5} \left[ 1-5 \left( c_i^2+c_j^2 \right) -15 c_i^2 c_j^2 + 2 \left( s_i s_j c - 4 c_i c_j \right)^2 \right],
\label{uQ}
\end{eqnarray}
with $c_k={\rm cos} \theta_k$, $s_k={\rm sin} \theta_k$, and $c={\rm cos} \phi_{ij}$. Herein ${\bm r}_{ij}$ is the centre-centre distance vector of two molecules $i$ and $j$, $r_{ab}$ is one of the four Lennard-Jones site-site distances; $a$ counts the two sites of molecule $i$, $b$ counts those of molecule $j$. The vectors ${\bm \omega}_i$ and ${\bm \omega}_j$ represent the orientations of the two molecules. $\theta_i$ is the angle between the axis of the molecule $i$ and the centre-centre connection line and $\phi_{ij}$ is the azimutal angle between the axis of molecule $i$ and $j$. The Lennard-Jones parameters $\sigma$ and $\epsilon$ represent size and energy respectively.

Among all possible arrangements of the charges in a quadrupole we have chosen the four charges to be arranged along the molecular axis in the symmetric sequence $+$, $-$, $-$, $+$ or, having the same energetic effect in pure fluids, $-$, $+$, $+$, $-$. In the large distance approximation the quadrupole interaction reduces to an ideal pointquadrupole interaction described by Eq. (\ref{uQ}). This reduces the number of parameters related to the quadrupole to one, namely the quadrupolar momentum $Q$.

Beyond a certain elongation, for small intermolecular distances $|\bm{r}_{ij}|$ the positive Len\-nard-Jones term $u_{\rm 2CLJ}$ of the pair potential cannot outweigh the divergence to $-\infty$ of the quadrupolar term $u_{\rm Q}$. This divergence of $u_{\rm 2CLJQ}$ leads to infinite Boltzmann factors, i.e. non-existence of the configurational integral. During molecular dynamics phase space sampling within the pressure range in question, this artefact of the 2CLJQ potential causes no problem as intermolecular centre-centre distances are very improbable to fall below critical values. However, the calculation of the chemical potential by test particle insertion often runs into critical intermolecular centre-centre distances. To avoid computational problems we followed the proposition of M\"oller et al. \cite{moeller9435} to put a hard sphere of diameter $0.4 \cdot \sigma$ directly on the quadrupole site to shield the quadrupolar interaction in critical cases. This hard sphere was not active during configuration generation.

The parameters $\sigma$ and $\epsilon$ of the 2CLJQ pair potential were used for the reduction of the thermodynamic properties as well as the model parameters $L$ and $Q^2$: $T^*=Tk/\epsilon$, $p^*=p\sigma^3/\epsilon$, $\rho^*=\rho\sigma^3$, $h^*=h/\epsilon$, $L^*=L/\sigma$, $Q^{*2}=Q^2/\left( \epsilon\sigma^5 \right)$, $\Delta t^*=\Delta t\sqrt{m/\epsilon}/\sigma$.

The reduced parameters $L^*$ and $Q^{*2}$ were varied in this investigation: $L^*=0$; $0.2$; $0.4$; $0.505$; $0.6$; $0.8$ and $Q^{*2}=0$; $1$; $2$; $3$; $4$. Combining these values leads to a set of 30 model fluids to be investigated here.

In order to achieve a monotonous transition from $L^*>0$ to $L^*=0$ we treated the spherical fluids 1CLJ, where $Q^{*2}=0$, and 1CLJQ as two centre LJ fluids with $L^*=0$. This leads to site superposition that is not present  when 1CLJ and 1CLJQ fluids are represented by a simple LJ site. Therefore the reduced temperatures, reduced pressures, reduced enthalpies and reduced quadrupolar momenta here are fourfold of the corresponding values in the one site case. Densities, of course, are not concerned.

For all simulations the centre-centre cut-off radius $r_{\rm c}$ was set to $5.0 \cdot \sigma$. Outside the cut-off sphere the fluid was assumed to have no preferential relative orientations of the molecules, i.e., in the calculation of the long range corrections, orientational averaging was done with equally weighted relative orientations as proposed by Lustig \cite{lustig8817}. The quadrupolar interaction needs no long range correction as it disappears by orientational averaging.

\section{Molecular simulation method for VLE data}
{\label{MSMFVD}}

For all 30 model fluids the $N\!pT$+Test Particle Method ($N\!pT$+TP Method) proposed by M\"oller et al. \cite{moeller9046, moeller9214} was applied to obtain the VLE data. A detailed description of this method has been given by M\"oller et al. \cite{moeller9435}, so we only sketch the key ideas here. The $N\!pT$+TP Method performs separate $N\!pT$ simulations in the liquid and the vapour phase and uses information about the pressures and the chemical potentials to calculate the VLE. Vapour pressures, saturated densities and residual enthalpies $h^{\mbox{\scriptsize res*}} \left( T^*,\rho^* \right) =h^* \left( T^*,\rho^* \right)-h^{\mbox{\scriptsize id*}} \left( T^* \right)$ are evaluated. Test particle insertion for the calculation of the residual part of the chemical potential in the $N\!pT$ ensemble is based on Widom's method \cite{widom6328}. Configuration space sampling was done by $N\!pT$-molecular dynamics simulations with $N=864$ particles for both liquid and vapour simulations. The dimensionless integration time step was set to $\Delta t^*=0.0015$. Starting from a face centred lattice arrangement every simulation run was given $10,000$ integration time steps to equilibrate. Data production was performed over $n=100,000$ integration time steps. At each production time step the information from insertion of $2N$ test particles in the liquid, and $N$ test particles in the vapour was used to calculate the chemical potential. The dimensionless dynamical parameter of $N\!pT$-MD-simulations ascribed to the box membrane was set to $2 \cdot 10^{-4}$ for liquid simulations and to $10^{-6}$ for vapour simulations. The high value of $N$ allowed simulations in the neighborhood of the critical point, and the high value of $n$ was used in order to obtain small statistical uncertainties.

On the basis of the contributions to the free energy $A$ of the 2CLJ interaction, i.e. $A_{\rm 2CLJ}$ from Mecke et al. \cite{mecke9768}, and of the quadrupolar interaction $A_{\rm Q}$ from Saager et al. \cite{saager9267}, one can construct $A_{\rm 2CLJQ}=A_{\rm 2CLJ}+A_{\rm Q}$. This construction of a hybrid equation of state (2CLJQ EOS) is analogous to the method applied to the two centre Lennard-Jones plus pointdipole fluid (2CLJD) by Kriebel et al. \cite{kriebel9815}. In planning our investigation we used this 2CLJQ EOS to estimate the critical temperatures $T^*_{\rm c,EOS} \left( Q^{*2},L^* \right)$ for the systems considered. VLE data were determined for temperatures $0.55$, $0.60$, $0.65$, $0.70$, $0.75$, $0.80$, $0.85$, $0.90$, $0.925$, $0.95 \cdot T^*_{\rm c,EOS}(Q^{*2},L^*)$.

Liquid simulations were performed for the whole temperature range, whereas vapour simulations were only performed at temperatures above $0.80 \cdot T^*_{\rm c,EOS}$. Below this temperature the second virial coefficient is sufficient for the VLE calculations. Liquid simulations took about eleven hours, vapour simulations about three hours CPU time on a modern workstation (e.g. Compaq AlphaStation XP1000).

\section{Simulation results of VLE data and critical data}

Table \ref{t2cljqvle} reports an extract of the VLE data of all 30 model fluids. Given are the vapour pressure $p_{\sigma}^*$, the saturated liquid density $\rho'^*$, the saturated vapour density $\rho''^*$, the residual saturated liquid enthalpy $h'^{\mbox{\scriptsize res*}}$, and the residual saturated vapour enthalpy $h''^{\mbox{\scriptsize res*}}$ for temperatures $T^* \approx 0.55 \cdot T^*_{\rm c}$, $0.80 \cdot T^*_{\rm c}$, and $0.95 \cdot T^*_{\rm c}$. Statistical uncertainties were determined with the method of Fincham et al. \cite{fincham8645} and the error propagation law.

\begin{figure}[ht]
\epsfig{file=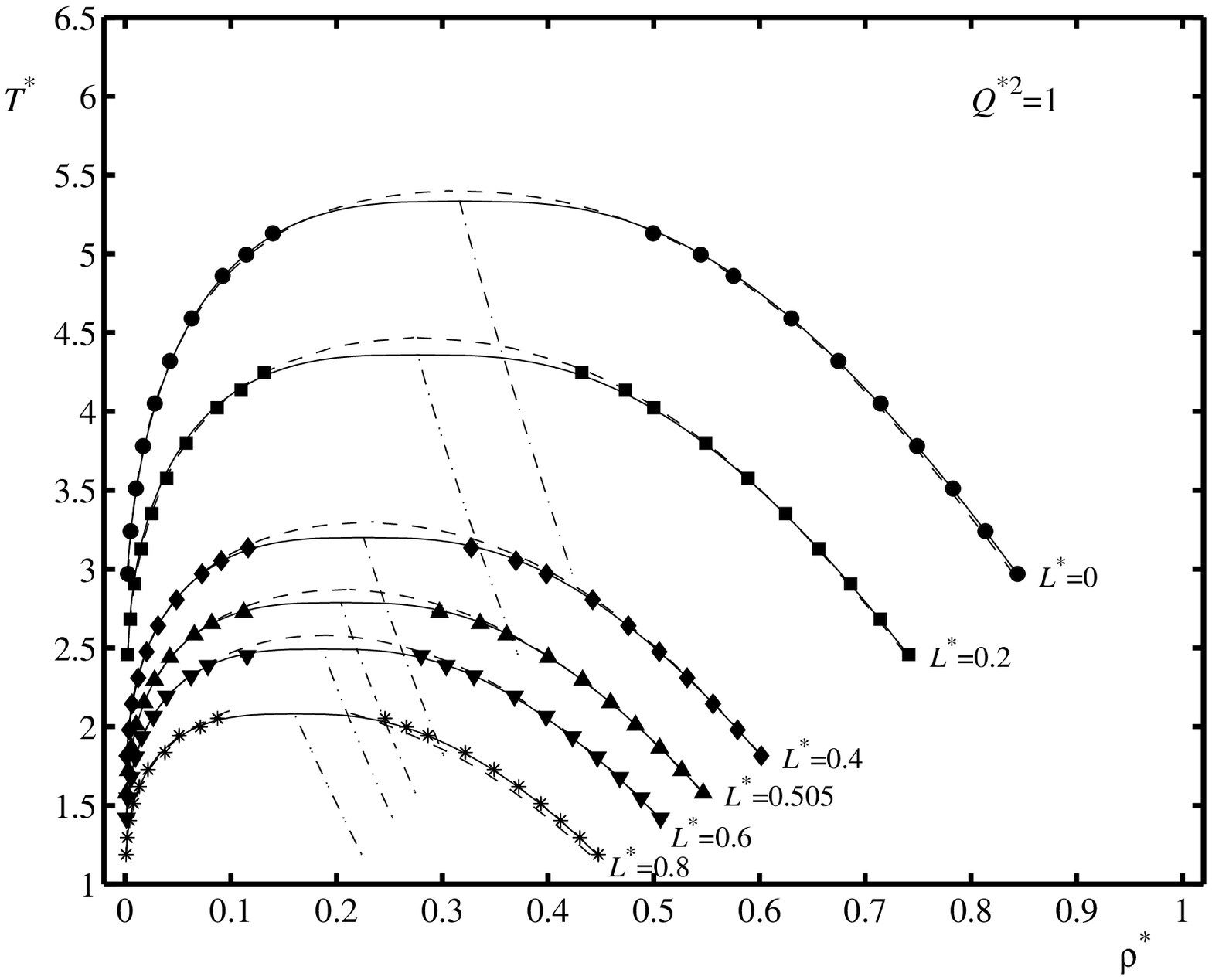,scale=1,angle=90}  
\caption[Temperature--density coexistence curves of the 2CLJQ model fluids with $Q^{*2}=1$. Symbols: simulation data. Full lines are values from the present correlation. Dashed-pointed lines represent the law of rectilinear diameters from the present correlation. Dashed lines are values from the 2CLJQ EOS \cite{mecke9768,saager9267}. All errorbars are within symbol size.]{}
\label{xa2Q1}
\end{figure}

\begin{figure}[ht]
\epsfig{file=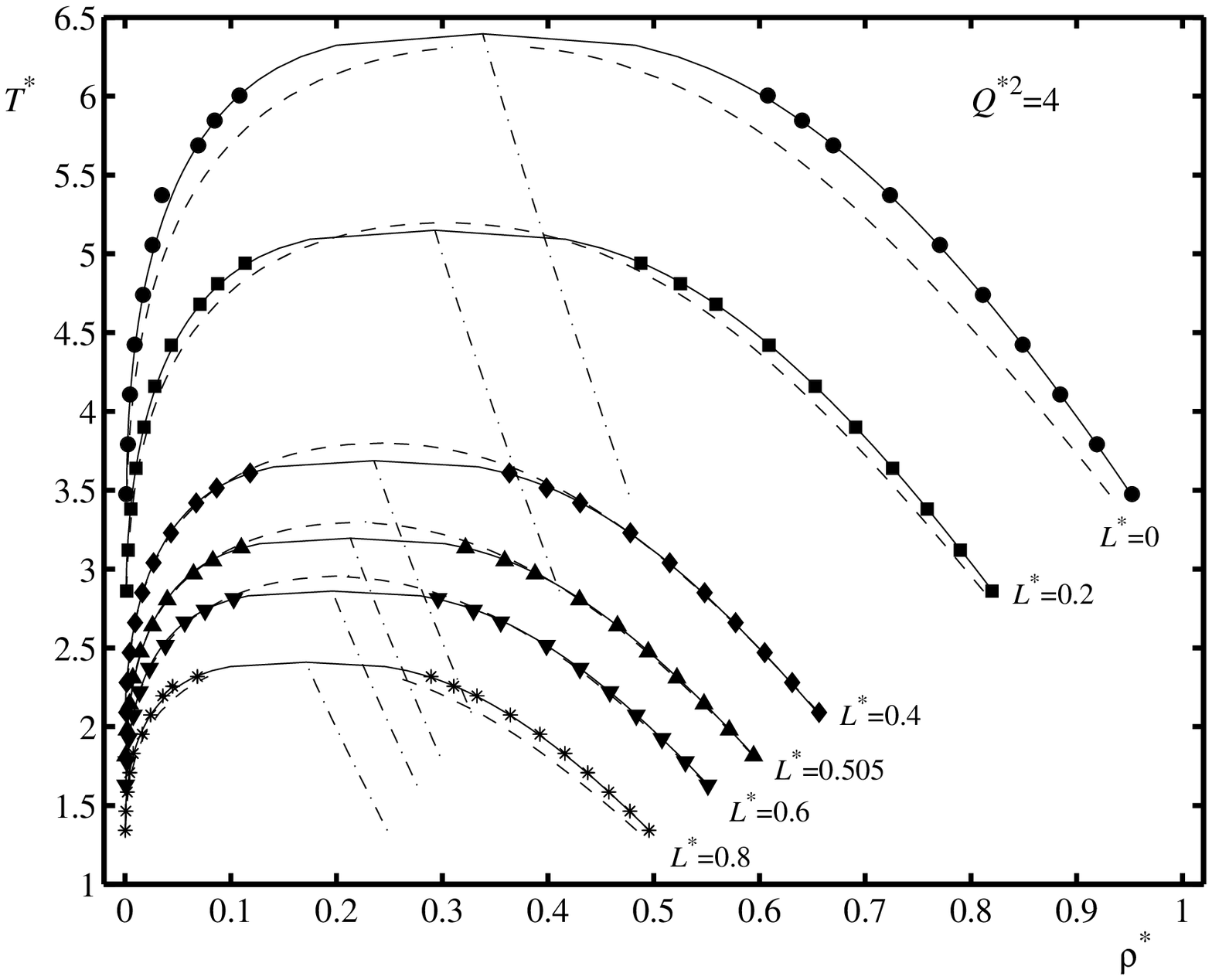,scale=1,angle=90}  
\caption[Temperature--density coexistence curves of the 2CLJQ model fluids with $Q^{*2}=4$. Symbols: simulation data. Full lines are values from the present correlation. Dashed-pointed lines represent the law of rectilinear diameters from the present correlation. Dashed lines are values from the 2CLJQ EOS \cite{mecke9768,saager9267}. All errorbars are within symbol size.]{}
\label{xa2Q4}
\end{figure}

\begin{figure}[ht]
\epsfig{file=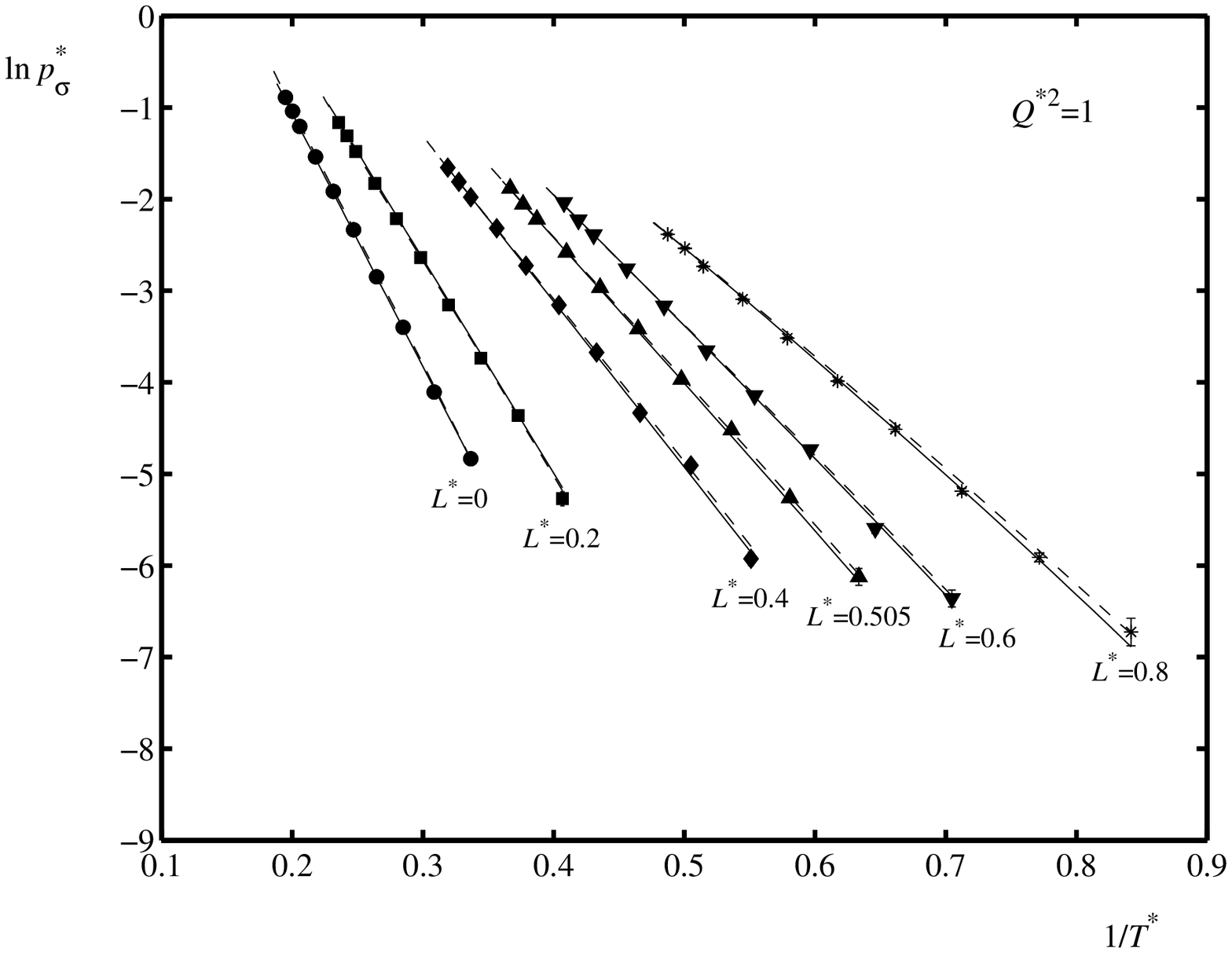,scale=1,angle=90}  
\caption[Vapour pressure curves of the 2CLJQ model fluids with $Q^{*2}=1$. Symbols: simulation data. Full lines are values from the present correlation. Dashed lines are values from the 2CLJQ EOS \cite{mecke9768,saager9267}.]{}
\label{xa1Q1}
\end{figure}

\begin{figure}[ht]
\epsfig{file=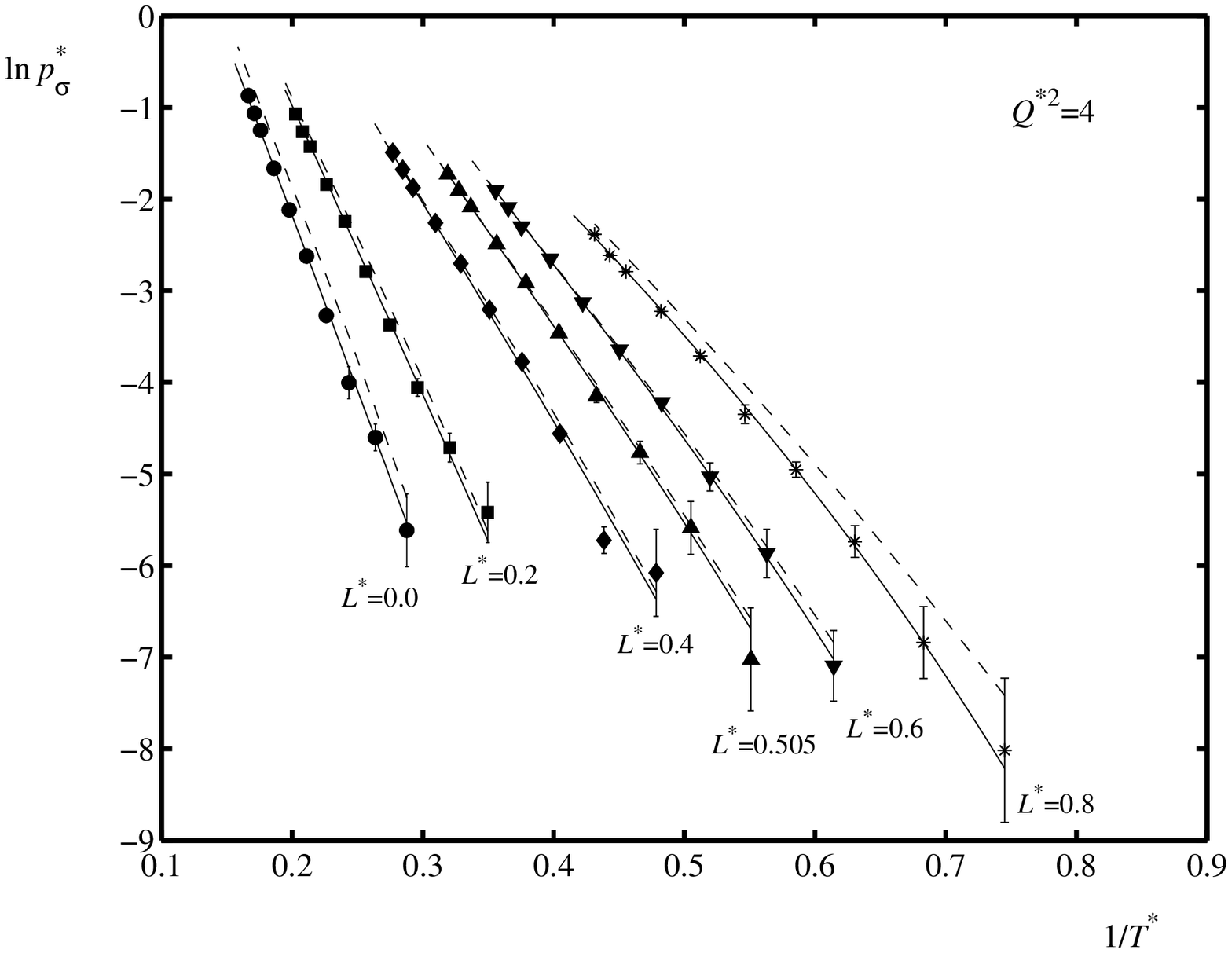,scale=1,angle=90}  
\caption[Vapour pressure curves of the 2CLJQ model fluids with $Q^{*2}=4$. Symbols: simulation data. Full lines are values from the present correlation. Dashed lines are values from the 2CLJQ EOS \cite{mecke9768,saager9267}.]{}
\label{xa1Q4}
\end{figure}

Figs. \ref{xa2Q1}, \ref{xa2Q4}, \ref{xa1Q1}, \ref{xa1Q4} illustrate for $Q^{*2}=1$ and $Q^{*2}=4$ the strong influence of both the elongation and the quadrupolar momentum on the 2CLJQ VLE data. Increasing the elongation or increasing the quadrupolar momentum strongly influences the shape of the density coexistence curve and the slope of the vapour pressure curve.

When the present 2CLJQ VLE data are compared to the 2CLJQ EOS as described in section \ref{MSMFVD}, systematic deviations are observed. This can best be seen in Figs. \ref{xa2Q4} and \ref{xa1Q4}. For models with small quadrupolar momenta and for models with elongations near $L^*=0.505$ we find good agreement. This is due to the fact, that the quadrupolar contribution to the EOS is based on data of a 2CLJQ model fluid with $L^*=0.505$. However outside the range mentioned above considerable deviations between the EOS and present 2CLJQ VLE data are observed. In most cases the EOS underestimates the saturated liquid densities and overestimates the saturated vapour densities. Where significant deviations occur, the vapour pressures from the 2CLJQ EOS are higher than those from the present study.

In order to determine the critical data we used the method of Lotfi et al. \cite{lotfi9213}, who, by simple means, found reliable critical data for the 1CLJ model fluid. Their values $T^*_{\rm c}=1.310$ and $\rho^*_{\rm c}=0.314$ are very close to those indicated later by Potoff et al. \cite{potoff9810}. It is known that the density--temperature dependence near the critical point is well described by $\rho^* \sim \left( T^*_{\rm c}-T^*\right)^{1/3}$, as given by Guggenheim \cite{guggenheim4525}, \cite{rowlinson1969}. In order to correlate the saturated densities we followed Lotfi et al. \cite{lotfi9213}
\begin{eqnarray}
\rho'^*=\rho^*_{\rm c} + C_1 \cdot (T^*_{\rm c}-T^* )^{1/3} + C'_2 \cdot (T^*_{\rm c}-T^*) + C'_3 \cdot (T^*_{\rm c}-T^*)^{3/2},
\label{Trho1corrcore} \\
\rho''^*=\rho^*_{\rm c} - C_1 \cdot (T^*_{\rm c}-T^* )^{1/3} + C''_2 \cdot (T^*_{\rm c}-T^*) + C''_3 \cdot (T^*_{\rm c}-T^*)^{3/2}.
\label{Trho2corrcore}
\end{eqnarray}
The simultaneous fitting of saturated liquid and saturated vapour densities yields not only the coefficients $C_1$, $C'_2$, $C'_3$, $C''_2$, $C''_3$, but also the critical data $\rho^*_{\rm c}$, $T^*_{\rm c}$. The critical temperatures and densities for the 2CLJQ model fluids are listed in Table \ref{2CLJQcdomtab}. In most cases present critical temperatures are lower than those estimated by the 2CLJQ EOS. Table \ref{2CLJQcdomtab} also contains the critical compressibility factor $Z_{\rm c}=p^*_{\rm c}/\left(\rho^*_{\rm c} T^*_{\rm c}\right)$. $Z_{\rm c}$ being a non-reduced property of the 2CLJQ fluids, it is of particular interest for comparisons to real quadrupolar fluids.

We estimated the uncertainties of $T^*_{\rm c}$ and $\rho^*_{\rm c}$ by refitting VLE data sets that had been perturbed stochastically within their own uncertainties. To span the whole range of the parameters $Q^{*2}$ and $L^*$, we have chosen the model fluids $Q^{*2}=0$, $L^*=0$ and $Q^{*2}=4$, $L^*=0.8$ for these estimations. Ten perturbed VLE data sets for both model fluids have been generated. Standard deviations were about $\sigma\!\left( T_{\rm c} \right) \approx 0.005$ and $\sigma\!\left( \rho_{\rm c} \right) \approx 0.0005$. From these results we conclude, that the critical temperatures calculated by this method are certain up to the second, the critical densities up to the third digit after the decimal point.

\clearpage

\section{Global correlation of VLE data}

In order to obtain VLE data for the whole range of $Q^{*2}$, $L^*$ and $T^*$ we have globally correlated the present data. We consider the critical data $T^*_{\rm c}(Q^{*2},L^*)$, $\rho^*_{\rm c}(Q^{*2},L^*)$, the saturated liquid density $\rho'^*(Q^{*2},L^*,T^*)$ and the vapour pressure $\ln p^*_{\sigma}(Q^{*2},L^*,T^*)$ to be the key VLE data for an adjustment to real fluids. The adequate shape of the temperature--density coexistence curve was achieved by simultaneously correlating the functions $\rho'^*\left(Q^{*2},L^*,T^*\right)$ and  $\rho''^*\left(Q^{*2},L^*,T^*\right)$. We want to point out, that it was not in the scope of the present investigation to construct a new 2CLJQ EOS. The correlation developed here is not designed to compete with an EOS. It shall merely be a working tool for a restricted field of application, namely the adjustment of model parameters to data of real fluids. The vapour pressure correlation is also used to test the thermodynamic consistency of the VLE data from simulations by the means of the Clausius-Clapeyron equation. Moreover, the correlations are useful for comparisons with results of other investigators. Details of the correlation method are described in the Appendix which also contains the resulting correlation functions (cf. Table \ref{corrtab}).

\subsection{Critical properties}

\begin{figure}[ht]
\epsfig{file=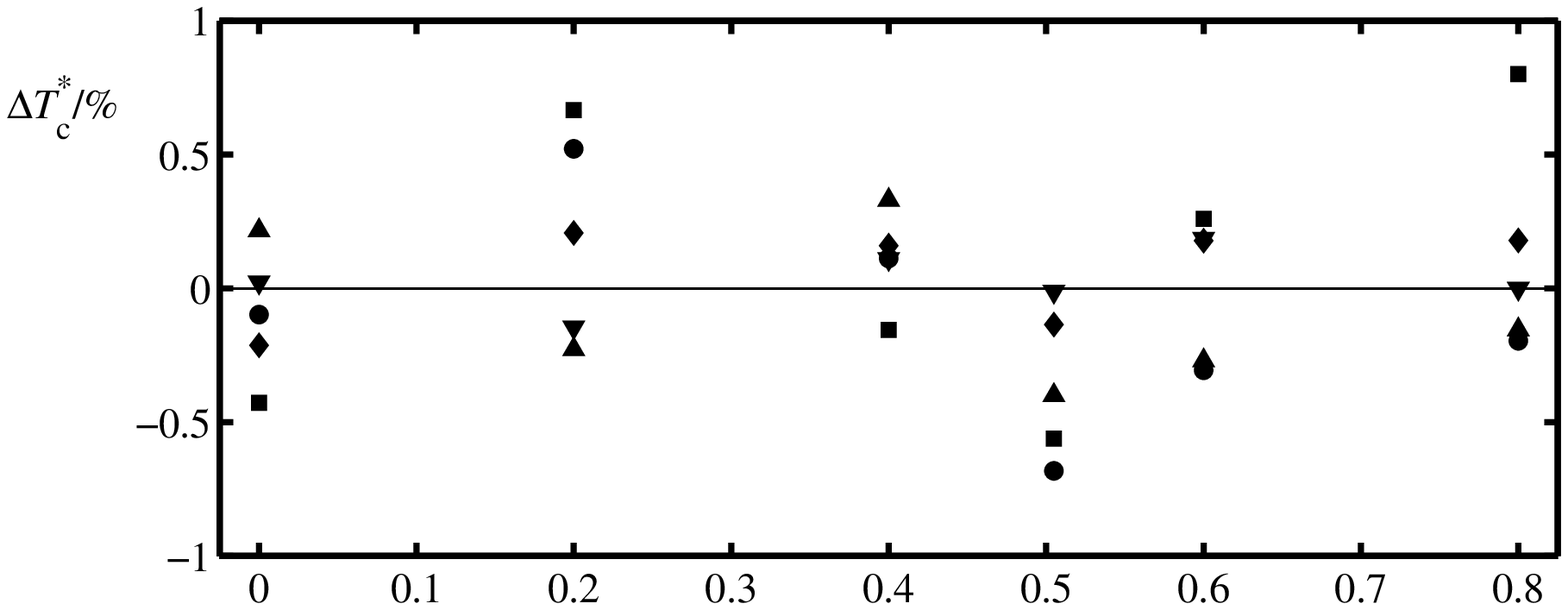,scale=1,angle=90}   
\epsfig{file=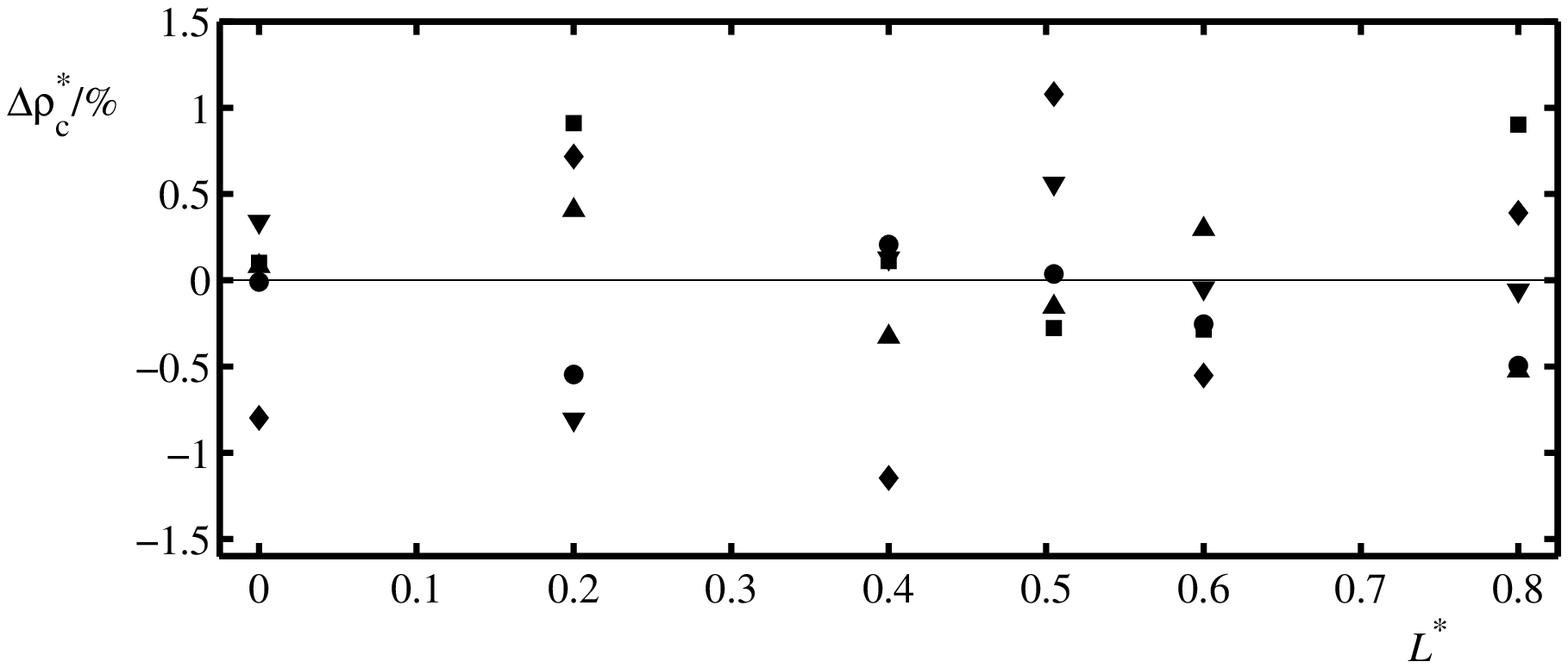,scale=1,angle=90} 
\caption[Relative deviations of simulation data to the correlation. Top: critical temperatures $\Delta T^*_{\rm c}=(T^*_{\rm c,sim}-T^*_{\rm c,corr})/T^*_{\rm c,corr}$. Bottom: critical densities $\Delta \rho^*_{\rm c}=(\rho^*_{\rm c,sim}-\rho^*_{\rm c,corr})/\rho^*_{\rm c,corr}$. Symbols are: $Q^{*2}=0$ ({\large $\bullet$}), $1$ ({\Huge $\centerdot$}), $2$ ({\footnotesize $\blacklozenge$}), $3$ ($\blacktriangle$), $4$ ($\blacktriangledown$).]{}
\label{xrTcRhoc}
\end{figure}

The correlation functions $T^*_{\rm c}(Q^{*2},L^*)$ and $\rho^*_{\rm c}(Q^{*2},L^*)$ were assumed to be linear combinations of elementary functions. The elementary functions and their coefficients are given in Table \ref{corrtab} in the Appendix. The quality of the correlations can be studied in Fig. \ref{xrTcRhoc}. Most relative deviations of $T^*_{\rm c}$ are within 0.5 \% which is only slightly more than the individual uncertainty estimated before. The critical densities are represented with roughly the same quality. It should be mentioned that possible systematic errors may be introduced to the critical data by the choice of the exponent $1/3$ in the second term of Eqs. (\ref{Trho1corrcore}) and (\ref{Trho2corrcore}). Both correlations are strictly monotonous in $Q^{*2}$ and $L^*$.

\subsection{Saturated densities, vapour pressure}

The temperature--density correlation is based on Eqs. (\ref{Trho1corrcore}) and (\ref{Trho2corrcore}), which have five adjustable parameters $C_1$, $C'_2$, $C'_3$, $C''_2$, and $C''_3$. Another three parameters are introduced by the correlation of the vapour pressure (cf. Appendix). Again the functions describing the dependency of the correlation parameters on $Q^{*2}$ and $L^*$ were assumed to be linear combinations of elementary functions. The elementary functions and their coefficients are given in Table \ref{corrtab} in the Appendix.

\begin{figure}[ht]
\epsfig{file=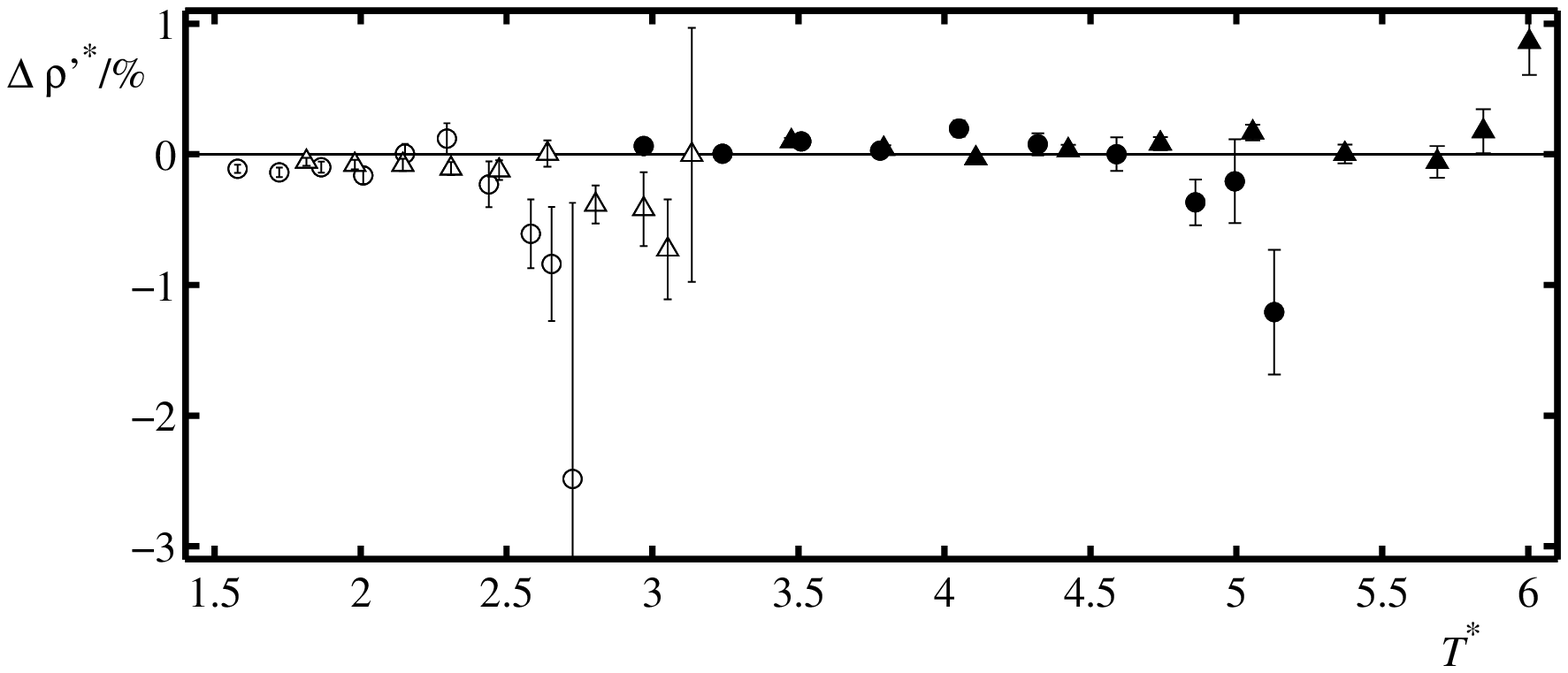,scale=1,angle=90}   
\epsfig{file=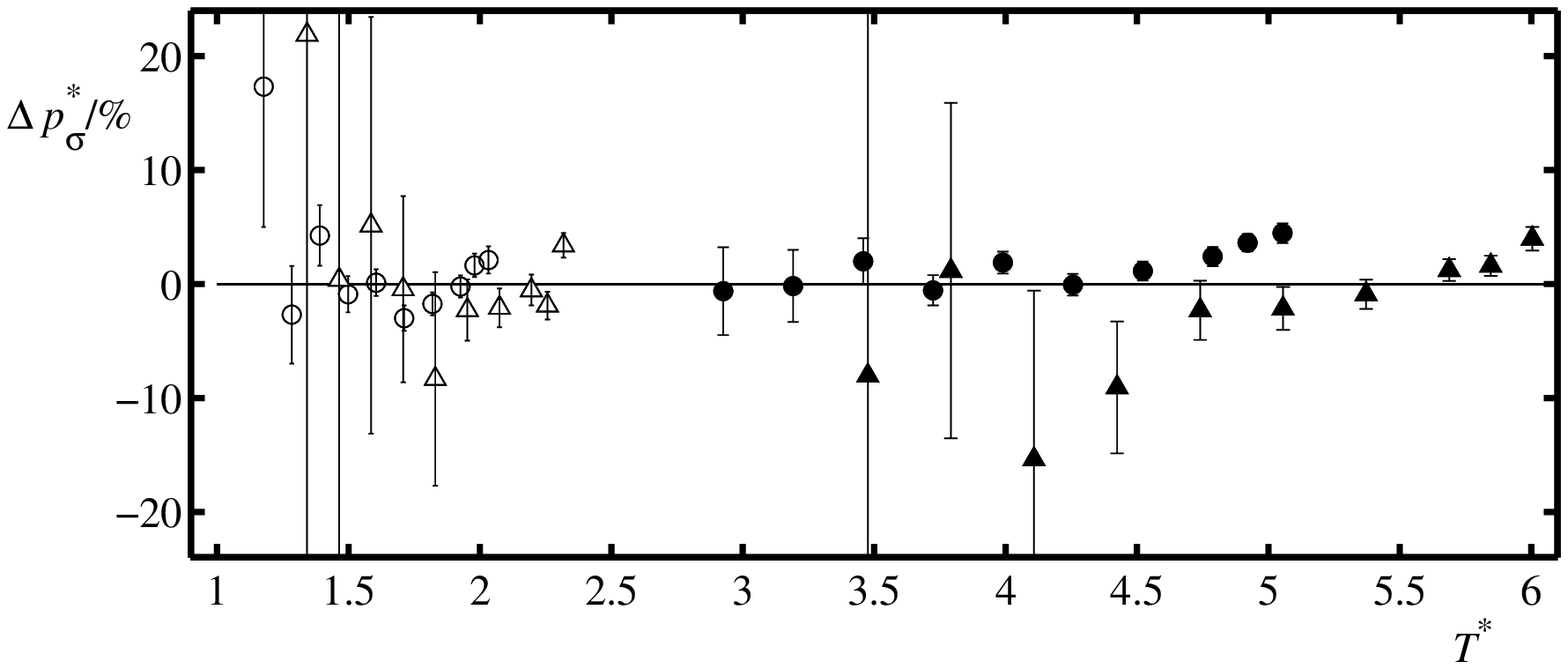,scale=1,angle=90}   
\caption[Relative deviations of simulation data to the correlation. Top: saturated liquid densities $\Delta \rho'^*=(\rho'^*_{\rm sim}-\rho'^*_{\rm corr})/\rho'^*_{\rm corr}$. Full and blank symbols are for $L^*=0$ and $L^*=0.505$ respectively, with $Q^{*2}=1$ ({\large $\bullet$}, $\circ$), $4$ ($\blacktriangle$, $\triangle$). Bottom: vapour pressures $\Delta p^*_{\rm \sigma}=(p^*_{\sigma \rm,sim}-p^*_{\sigma,\rm corr})/p^*_{\sigma,\rm corr}$. Full and blank symbols are for $L^*=0$ and $L^*=0.8$ respectively, with $Q^{*2}=0$ ({\large $\bullet$}, $\circ$), $4$ ($\blacktriangle$, $\triangle$).]{}
\label{xrRlpsQLT_someL_someQ}
\end{figure}

A comparison between the correlations and the simulation data can be seen in Figs. \ref{xa2Q1}, \ref{xa2Q4}, \ref{xa1Q1}, and \ref{xa1Q4}.
A more detailed comparison is given in Fig. \ref{xrRlpsQLT_someL_someQ}. Exemplarily, the relative deviations of the simulation results from the correlations are shown. Typically, the saturated liquid density correlation has the largest deviations for the highest temperature of each model, which is due to the large uncertainties in the near critical region. The correlation is smooth and shows small deviations in the range of 0.4 \% for mid temperatures which are most important for adjustments to real fluids.

Relative deviations of the saturated vapour densities are not illustrated here. The vapour density correlation should not be used below $0.60 \cdot T_{\rm c}(Q^{*2},L^*)$. At low temperatures the correlation is not useful as it does not capture the limiting case of the ideal gas, which is independent of the parameters $Q^{*2}$ and $L^*$.

In most cases, the vapour pressure correlation represents the simulation data within their uncertainties. It has to be mentioned, that the vapour pressures from simulation at low temperatures show high uncertainties due to the uncertain values of the chemical potential obtained by Widom's test particle insertion in dense liquid phases.

\begin{figure}[ht]
\epsfig{file=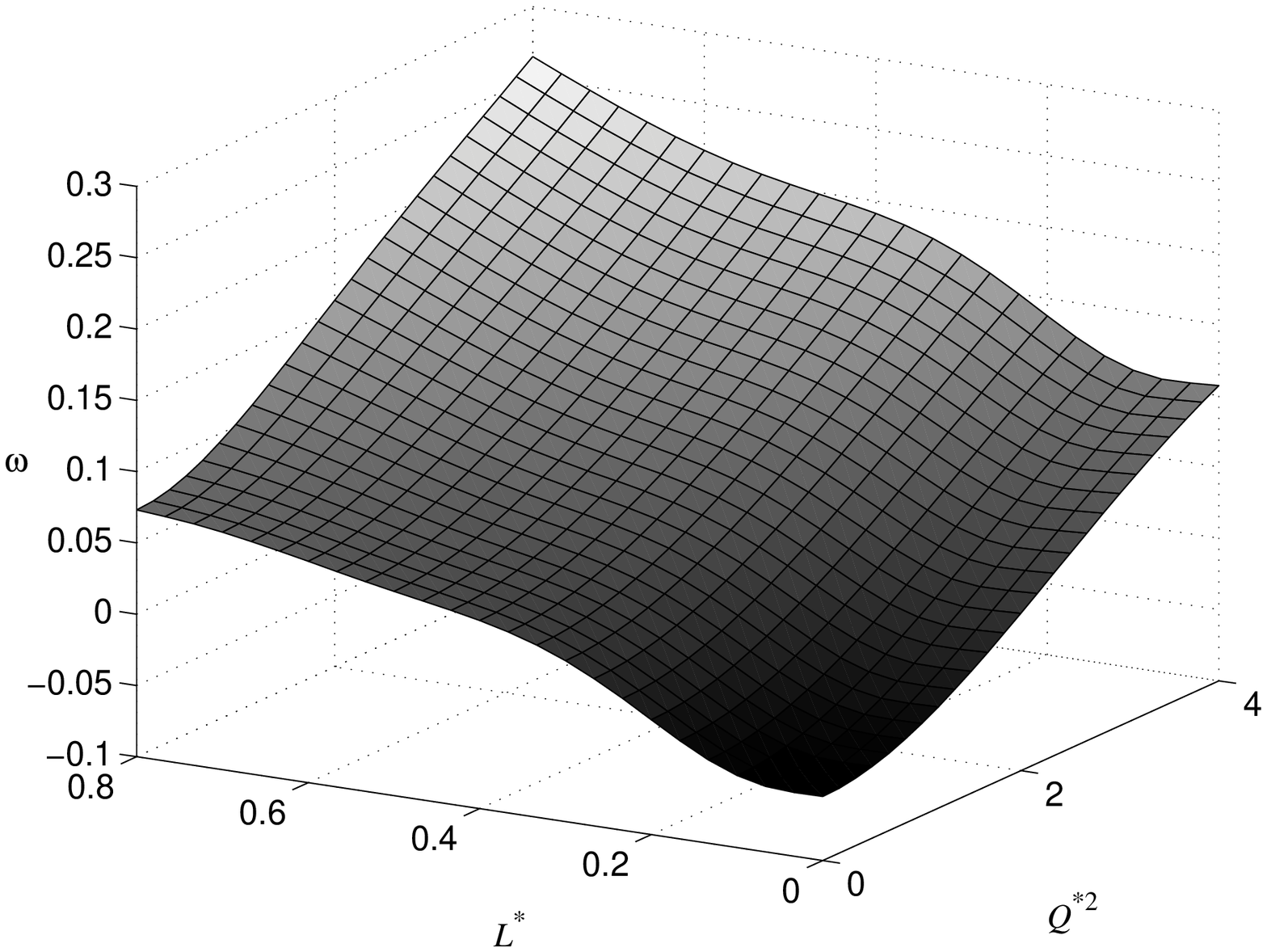,scale=1,angle=90}
\caption[The acentric factor $\omega$ of the 2CLJQ fluid vs. $Q^{*2}$ and $L^*$. Resulting from correlation from the present work.]{}
\label{xacf}
\end{figure}

By extrapolating the vapour pressure correlation slightly to the critical point, the acentric factor \cite{pitzer5534}
\begin{eqnarray}
\omega \left( Q^{*2},L^* \right)=-{\rm log}_{10}\frac{p^* \left( Q^{*2},L^*,0.7 \cdot T^*_{\rm c}\right)}{p^*_{\rm c} \left( Q^{*2},L^* \right)}-1,
\label{omega}
\end{eqnarray}
can be calculated easily from the correlations discussed before, cf. Fig. \ref{xacf}. We find that $\omega$ increases monotonously with increasing $Q^{*2}$ and $L^*$. For high $Q^{*2}$ at approximately $L^*=0.05$ there is a weakly distinct local minimum which probably is an artefact of the correlation. Table \ref{2CLJQcdomtab} contains the critical pressures and the acentric factors for the 2CLJQ model fluids calculated on the basis of present correlations.

\clearpage

\section{Discussion}

\subsection{Comparison to results of other authors}

\begin{figure}[ht]
\epsfig{file=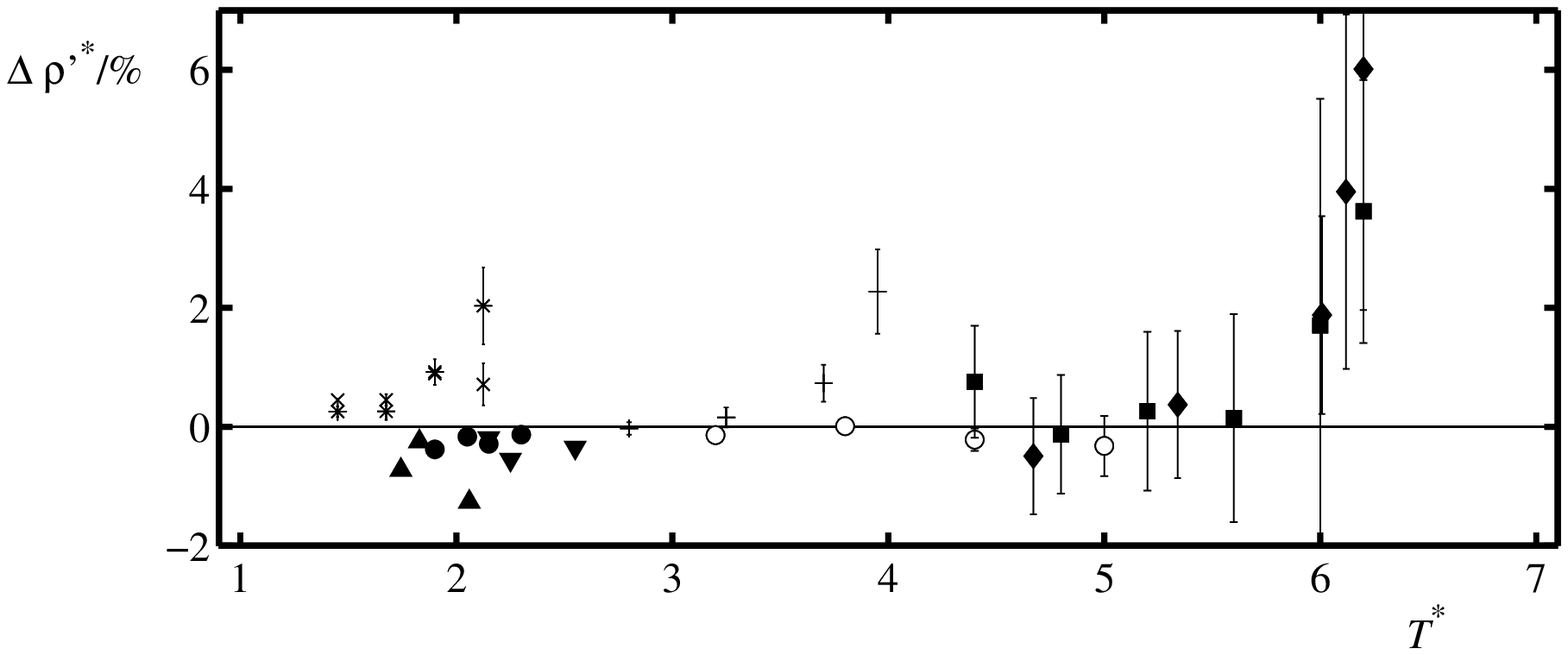,scale=1,angle=90}
\epsfig{file=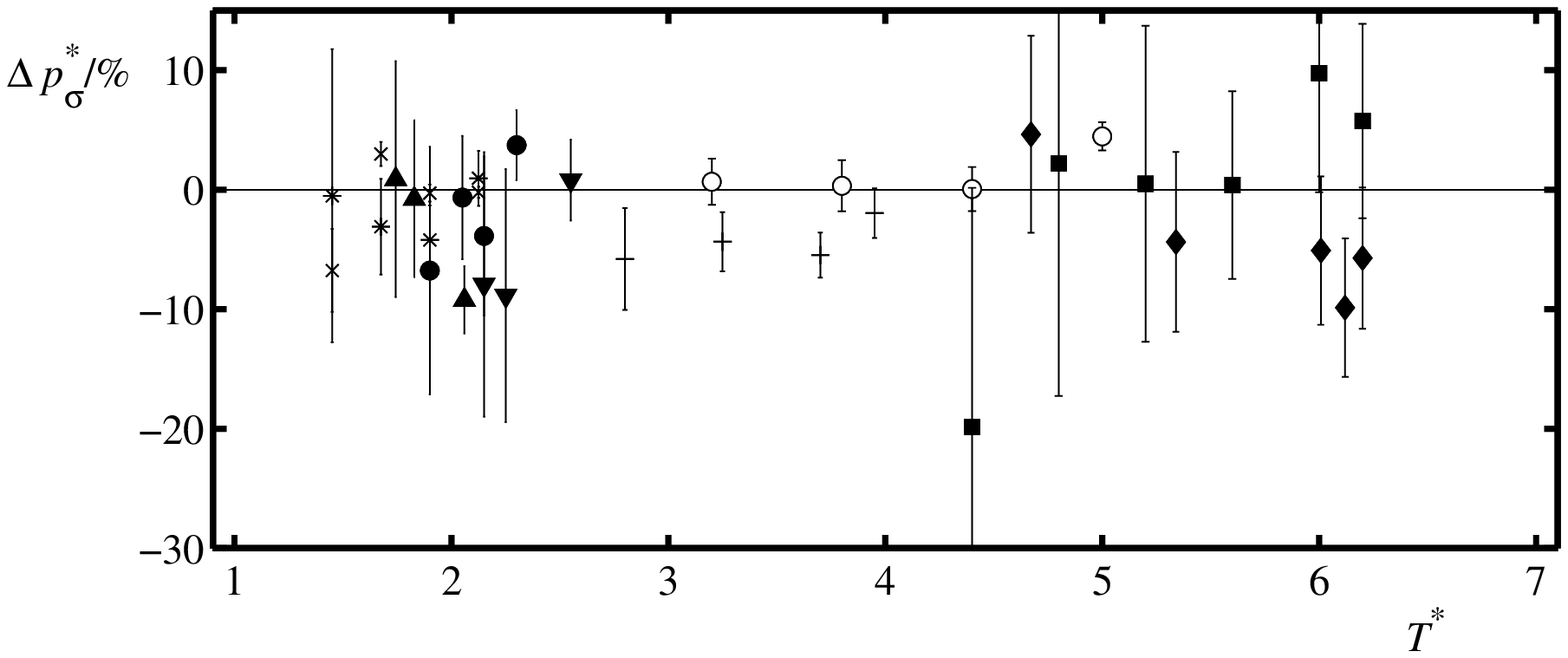,scale=1,angle=90}
\caption[Relative deviations of simulation data of other authors to the correlations based on simulations from the present work. Top: saturated liquid densities $\Delta \rho'^*=(\rho'^*_{\rm other}-\rho'^*_{\rm corr})/\rho'^*_{\rm corr}$. Bottom: vapour pressures $p^*_{\sigma}=(p^*_{\sigma, \rm other}-p^*_{\sigma, \rm corr})/p^*_{\sigma, \rm corr}$. Data are taken from: Lotfi et al. \cite{lotfi9213}, 1CLJ ($\circ$); Kriebel et al. \cite{kriebel9538}, 2CLJ with $L^*=0.22$ ($+$), 2CLJ with $L^*=0.67$ ($*$); Kronome et al. \cite{kronome9827}, 2CLJ with $L^*=0.67$ ($\times$); Stapleton et al. \cite{stapleton8914}, 1CLJQ with $Q^{*2}=4$ ({\footnotesize $\blacklozenge$}); Smit et al. \cite{smit9042}, 1CLJQ with $Q^{*2}=4$ ({\Huge $\centerdot$}); M\"oller et al. \cite{moeller9435}, 2CLJQ with $Q^{*2}=3.0255$ and $L^*=0.65$ ({\large $\bullet$}), 2CLJQ with $Q^{*2}=3.0255$ and $L^*=0.779$ ($\blacktriangle$), 2CLJQ with $Q^{*2}=3.7$ and $L^*=0.58$($\blacktriangledown$). Errorbars show the uncertainties, if they have been indi\-ca\-ted.]{}
\label{xvergl_1_3}
\end{figure}

Present results are compared with the simulation results of other authors. Fig. \ref{xvergl_1_3} presents the relative deviations for the saturated liquid density $\Delta \rho'^*=(\rho'^*_{\rm other}-\rho'^*_{\rm corr})/\rho'^*_{\rm corr}$ and the vapour pressure $\Delta p^*_{\sigma}=(p^*_{\sigma,\rm other}-p^*_{\sigma,\rm corr})/p^*_{\sigma,\rm corr}$. Other authors investigated subsystems of the 2CLJQ model fluid, Lotfi et al. the 1CLJ \cite{lotfi9213}, Kriebel et al. \cite{kriebel9538} and Kronome et al. \cite{kronome9827} the 2CLJ, Stapleton et al. \cite{stapleton8914}, Smit et al. \cite{smit9042} and O'Shea et al. \cite{oshea9723} the 1CLJQ, and M\"oller et al. \cite{moeller9435} the 2CLJQ. They applied different simulation methods: molecular dynamics $N\!pT$+TP Method, Monte Carlo $N\!pT$+TP Method, or Gibbs Ensemble Monte Carlo.

The saturated liquid densities of the other authors agree in almost all cases within the combined uncertainties of our correlation and of the simulation results. The near critical points of course show larger deviations. Also for the vapour pressure good agreement is observed.

Present 2CLJQ VLE data show lower statistical uncertainties than those of other authors, and they have the advantage that uniformly the same simulation method was used to produce them.

\subsection{Thermodynamic consistency test}

The thermodynamic consistency of the simulation data was checked with the Clausius-Clapeyron equation
\begin{eqnarray}
\frac{\partial {\rm ln}p^*_{\sigma}}{\partial T^*}=\frac{\Delta h^*_{\rm v}}{p^*_{\sigma} T^* \left( 1/\rho''^*-1/\rho'^* \right)}.
\label{clcl}
\end{eqnarray}
We used the present vapour pressure correlation in order to evaluate the left hand side (LHS) of Eq. (\ref{clcl}). The right hand side (RHS) of Eq. (\ref{clcl}) was calculated from our simulation data, the uncertainty was calculated by the error propagation law. The requirements of Eq. (\ref{clcl}) are fulfilled at almost all temperatures within the uncertainties of the RHS. We conclude that our data are thermodynamically consistent.

\subsection{Locus of the critical point}

The influence of the quadrupole $Q^{*2}$ and the elongation $L^*$ on the critical properties can be studied in Table \ref{2CLJQcdomtab} and in Figs. \ref{xa2Q1} and \ref{xa2Q4}. The critical temperature $T^*_{\rm c}$ and the critical density $\rho^*_{\rm c}$ increase together with $Q^{*2}$. The increase of $T^*_{\rm c}$ with $Q^{*2}$ is more important for molecules with small $L^*$ than for those with large $L^*$. However, this only holds for the absolute increase $\Delta T^*_{\rm c}$. The relative increase $\Delta T^*_{\rm c}/T^*_{\rm c}$ with increasing $Q^{*2}$ only weakly depends on $L^*$. This finding is in agreement with results from Garz\'on et al. \cite{garzon9441}, who studied the VLE of Kihara fluids with the same pointquadrupole we used. Garz\'on et al. predicted that the 2CLJQ model fluid would qualitatively behave in the same way as their Kihara plus pointquadrupole model fluid.

The critical pressure $p^*_{\rm c}$ increases with increasing $Q^{*2}$ for a fixed elongation $L^*$. The critical pressure $p^*_{\rm c}$ decreases strongly with increasing elongation $L^*$ for a fixed quadrupolar momentum $Q^{*2}$. The relative decrease $\Delta p^*_{\rm c}/p^*_{\rm c}$ with increasing $L^*$ only weakly depends on $Q^{*2}$. As shown in Figs. \ref{xa1Q1} and \ref{xa1Q4} the absolute value of the slope of the function $\ln p^*_{\sigma}$ vs. $1/T^*$, i.e. the enthalpy of vaporization $\Delta h^*_{\rm v}$, increases with increasing $Q^{*2}$ for a fixed elongation $L^*$. It decreases with increasing elongation $L^*$ for a fixed $Q^{*2}$. This is analogous to the influence of $Q^{*2}$ and $L^*$ on the locus of the critical pressure $p^*_{\rm c}$. It can also be seen that the vapour pressure $p^*_{\sigma}$ of a model fluid of given elongation $L^*$ decreases when the quadrupolar momentum $Q^{*2}$ is increased. These results harmonize with those of Garz\'on et al \cite{garzon9441}.

\subsection{Deviation from principle of corresponding states}

\begin{figure}[ht]
\epsfig{file=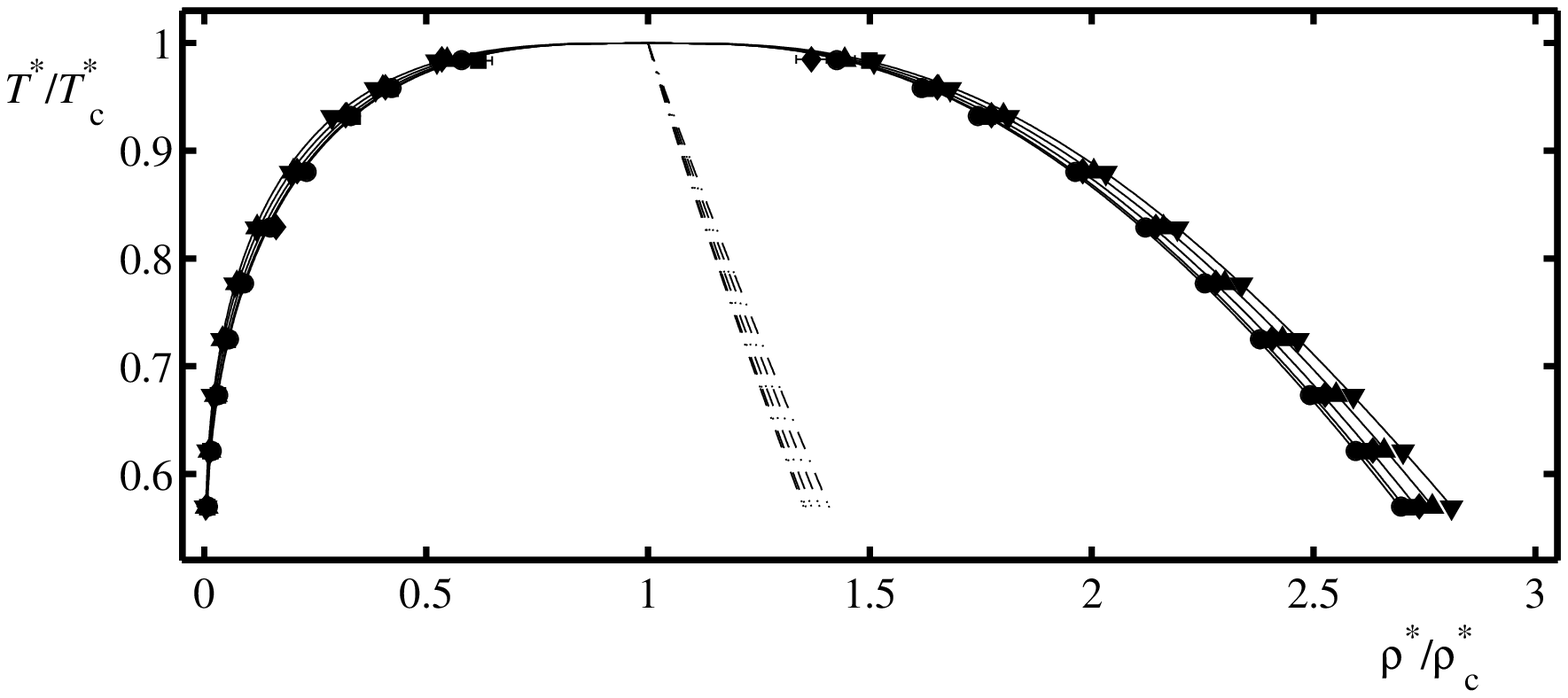,scale=1,angle=90}  
\epsfig{file=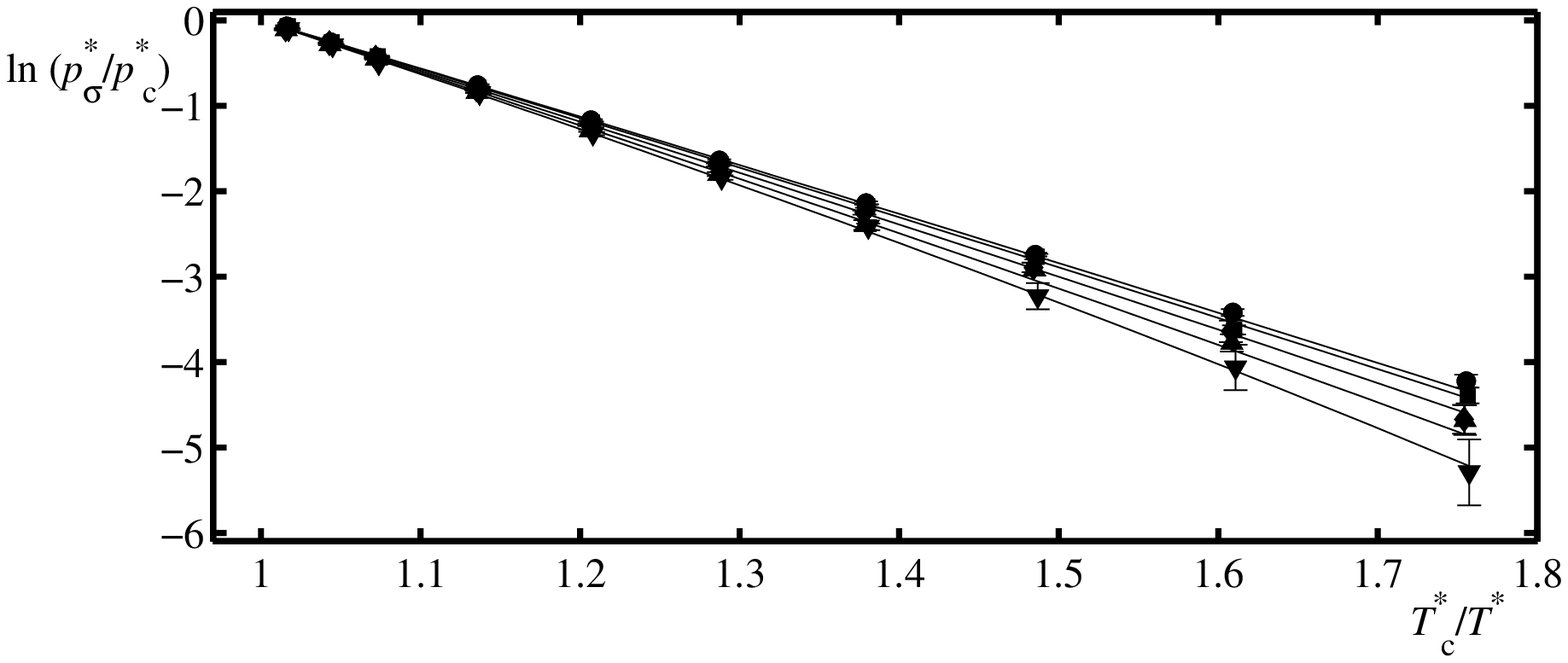,scale=1,angle=90}  
\caption[Deviation from the principle of corresponding states caused by the influence of the quadrupole. Top: saturated densities. Bottom: vapour pressures. Elongation is $L^*=0.6$, with $Q^{*2}=0$ ({\large $\bullet$}), $1$ ({\Huge $\centerdot$}), $2$ ({\footnotesize $\blacklozenge$}), $3$ ($\blacktriangle$), $4$ ($\blacktriangledown$).]{}
\label{xa2a1L06_ks}
\end{figure}

According to the principle of corresponding states for the 2CLJQ fluid, the curves $\ln \left(p^*_{\sigma}/p^*_{\rm c}\right)$ vs. $T^*_{\rm c}/T^*$ would form one unique curve regardless of the values of $Q^{*2}$ and $L^*$. This would also be true for the curves $T^*/T^*_{\rm c}$ vs. $\rho^*/\rho^*_{\rm c}$. However, molecular shape and presence of a quadrupole do cause deviations from the principle of corresponding states. The widening effect of the quadrupole on the density coexistence curve is shown in top of Fig. \ref{xa2a1L06_ks} for various 2CLJQ model fluids with $L^*=0.6$. Molecular anisotropy causes the same effect. This is a further confirmation for the conclusion of Garz\'on et al. \cite{garzon9441} that VLE coexistence density curves were broadened by the presence of multipoles. The displacement of the vapour pressure curves due to the quadrupole is presented in bottom of Fig. \ref{xa2a1L06_ks}. Qualitatively, the same effect is caused by molecular anisotropy. Fischer et al. \cite{fischer8448} have reported this effect as well.

These discernable deviations from the principle of corresponding states are reflected by the behaviour of the acentric factor $\omega$ vs. $Q^{*2}$ and $L^*$, cf. Eq. (\ref{omega}). The more a fluid deviates from the principle of corresponding states, the higher $\omega$ will be.

\begin{figure}[ht]
\epsfig{file=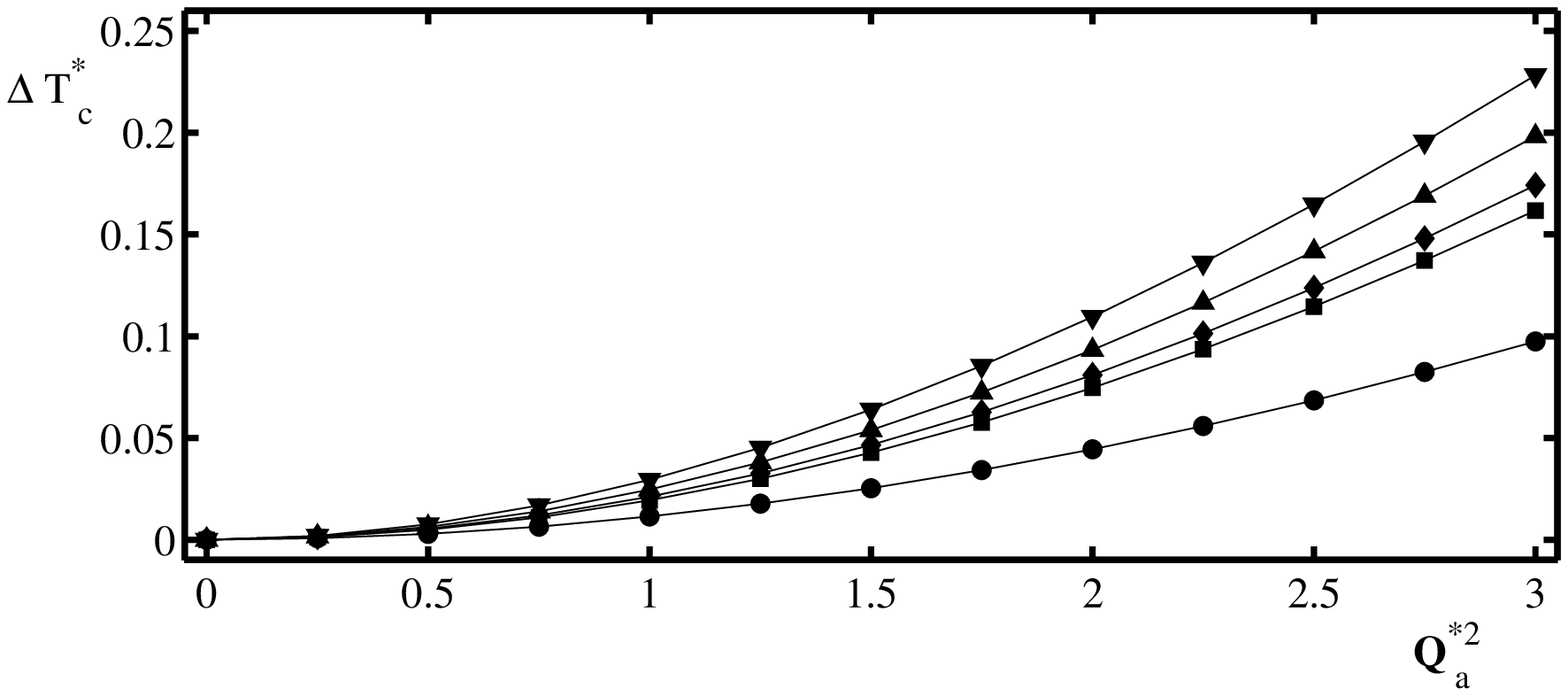,scale=1,angle=90}  
\epsfig{file=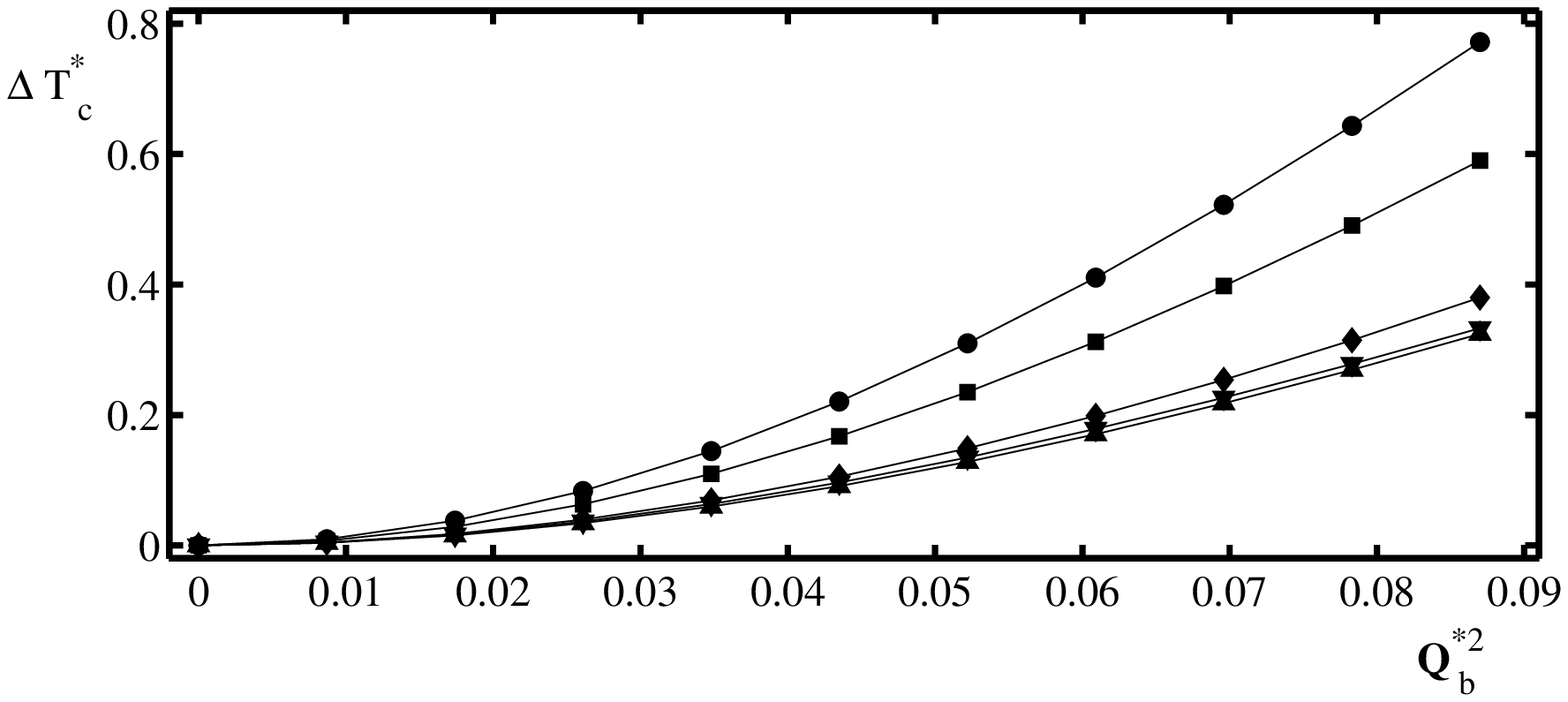,scale=1,angle=90}  
\caption[Deviation from principle of corresponding states for the increase of the critical temperature $\Delta T^*_{\rm c}$, cf. Eq. (\ref{DTc}). Top: $\Delta T^*_{\rm c}$ as a function of the ``reduced density of quadrupole'' ${\bm Q}^{*2}_{\rm a}$ and of the elongation $L^*$. Elongations are $L^*=0$ ({\large $\bullet$}), $0.2$ ({\Huge $\centerdot$}), $0.4$ ({\footnotesize $\blacklozenge$}), $0.6$ ($\blacktriangle$), $0.8$ ($\blacktriangledown$). Bottom: $\Delta T^*_{\rm c}$ as a function of the ``effective quadrupolar momentum'' ${\bm Q}^{*2}_{\rm b}$ and of the elongation $L^*$. Elongations are $L^*=0$ ({\large $\bullet$}), $0.2$ ({\Huge $\centerdot$}), $0.4$ ({\footnotesize $\blacklozenge$}), $0.6$ ($\blacktriangle$), $0.8$ ($\blacktriangledown$).]{}.
\label{xDTcQd}
\end{figure}

Garz\'on et al. \cite{garzon9441}, studying the quadrupolar Kihara fluid, introduced the ``reduced density of quadrupole'' ${\bm Q}^{*2}_{\rm a}$
\begin{eqnarray}
{\bm Q}^{*2}_{\rm a} \left( Q^{*2},L^* \right)=Q^{*2}/(V^{*5/3}_{\rm m} \left( L^* \right)),
\label{quaddens}
\end{eqnarray}
wherein
\begin{eqnarray}
V^*_{\rm m}\left( L^* \right)=\left( 1+1.5L^* \right) \cdot \pi/6 \nonumber
\end{eqnarray}
is the dimensionless volume of the Kihara molecule, in order to study the increase of the critical temperature
\begin{eqnarray}
\Delta T^*_{\rm c} \left( {\bm Q}^{*2}_{\rm a}\left( Q^{*2},L^* \right) \right)=T^*_{\rm c} \left( {\bm Q}^{*2}_{\rm a}\left( Q^{*2},L^* \right) \right) - T^*_{\rm c} \left( {\bm Q}^{*2}_{\rm a} \left( Q^{*2}=0,L^* \right) \right).
\label{DTc}
\end{eqnarray}
Within their statistical uncertainties they concluded that $\Delta T^*_{\rm c}$ vs. ${\bm Q}^2_{\rm a}$ forms a unique curve for fluids of different anisotropy, i.e. a principle of corresponding states. For the 2CLJQ fluid we introduce an analogous volume for two intersecting spheres with diameter $\sigma$ a distance $L^*$ apart
\begin{eqnarray}
V^*_{\rm m} \left( L^* \right)=\left(1+0.5 \left( 3L^*-L^{*3} \right)\right) \cdot \pi/6. \nonumber
\end{eqnarray}
Using it in Eq. (\ref{quaddens}) and calculating $\Delta T^*_{\rm c}$ according to Eq. (\ref{DTc}), we can not confirm that principle of corresponding states for the 2CLJQ fluid, cf. Fig. \ref{xDTcQd}, top. An alternative ``effective quadrupolar momentum'' suggested by Bohn et al. \cite{bohn8879}
\begin{eqnarray}
{\bm Q}^{*2}_{\rm b}=Q^{*2} \cdot \rho_{\rm c}^*\left(Q^{*2}=0,L^*\right)^{5/3}/T^*_{\rm c}\left(Q^{*2}=0,L^*\right).
\label{effquadred}
\end{eqnarray}
was examined in the same way. The pseudocritical data of unpolar fluids used by Bohn et al. \cite{bohn8879} have been replaced by present critical data of unpolar fluids. Fig. \ref{xDTcQd}, bottom, shows the application of Eq. (\ref{effquadred}) to Eq. (\ref{DTc}). Only for high elongations a better accordance to the principle of corresponding states can be observed.



\clearpage

\section{Conclusion}

The present paper is aimed at the qualitative and quantitative improvement of available VLE data of the 2CLJQ model fluid. In a systematic investigation the two parameters $Q^{*2}$ and $L^*$ of the model fluid have been varied in the ranges 0 to 4 and 0 to 0.8 respectively, studying 30 model fluids in detail, including the non-polar and spherical cases. The $N\!pT+TP$ Method has been applied for the production of VLE data in the temperature range from about $0.55 \cdot T^*_{\rm c}$ to $0.95 \cdot T^*_{\rm c}$ for all 30 model fluids. The comparison of data from the present work to a hybrid 2CLJQ EOS has shown the shortcomings of the EOS. Single fits of the saturated densities with Guggenheim's equations yielded critical data for all 30 systems.

In order to have useful tools for adjustments of dimensionless VLE data and critical data from the present work to real fluids we constructed global correlations of the critical data, the saturated density coexistence curve, and the vapour pressure curve. The correlations for saturated liquid densities and vapour pressures are sufficiently precise for mid-temperatures, which we consider the most important region for the intended adjustments. Data from the present work show no systematic deviations to results of other investigators. In many cases they how ever have lower uncertainties, furthermore the whole data set is self-consistent. The Clausius-Clapeyron equation has proven thermodynamic consistency of present data.

Using the correlations from the present work, we studied the influence of $Q^{*2}$ and $L^*$ on the locus of the critical point and on the shapes of the saturated density coexistence curves and of the vapour pressure curves. A deviation from the principle of corresponding states is found, due to the presence of polarity and anisotropy. Our results support those of Garz\'on et al \cite{garzon9441}. However, the principle of corresponding states for the increase of the critical temperature of fluids of different anisotropy but equal equivalent quadrupolar momentum can not be validated for the 2CLJQ model fluid neither with the ``reduced density of quadrupole'' suggested by Garz\'on et al. \cite{garzon9441} nor with the ``effective quadrupolar momentum'' of Bohn et al. \cite{bohn8879}.

The results from the present work pave the way to applications of molecular simulation data to the description of VLE of real quadrupolar fluids.

\section{List of Symbols}

\begin{tabular}{ll}
$A$ & coefficient of correlation function \\
$A$ & free energy \\
$a$ & interaction site counting index \\
$B$ & coefficient of correlation function \\
$b$ & interaction site counting index \\
$C$ & coefficient of correlation function \\
$c$ & coefficient of correlation function \\
$c$ & constant in set of elementary functions \\
$F$ & function to minimize \\
$G$ & function to minimize \\
$h$ & enthalpy \\
$i$ & data point counting index \\
$i$ & elementary function counting index \\
$i$ & particle counting index \\
$j$ & elementary function counting index \\
$j$ & particle counting index \\
$k$ & Boltzmann constant \\
$k$ & elementary function counting index \\
$k$ & particle counting index \\
$L$ & molecular elongation \\
LHS & left hand side of Clausius-Clapeyron equation \\
$\ell$ & simplified notation for $L^*$ \\
$m$ & mass of particle \\
$N$ & number of particles \\
$n$ & number of time steps \\
$p$ & pressure \\
$Q$ & quadrupolar momentum \\
${\bm Q}_{\rm a}$ & ``reduced density of quadrupole'' \\
${\bm Q}_{\rm b}$ & ``effective quadrupolar momentum'' \\
$q$ & simplified notation for $Q^{*2}$\\
RHS & right hand side of Clausius-Clapeyron equation \\
$r$ & site-site distance \\
$r_{\rm c}$ & centre-centre cut-off radius \\
$t$ & time \\
\end{tabular}
\clearpage

\noindent
\begin{tabular}{ll}
$T$ & temperature \\
$u$ & inner energy \\
$u$ & pair potential \\
$V_{\rm m}$ & molecular volume \\
$v$ & volume \\
$y$ & linear combination of elementary functions \\
$Z$ & compressibility factor \\
$\alpha$ & coefficient of elementary function \\
$\beta$ & coefficient of elementary function \\
$\gamma$ & coefficient of elementary function \\
$\Delta$ & relative deviation \\
$\Delta CC$ & relative deviation of RHS from LHS \\
$\Delta h_{\rm v}$ & enthalpy of vaporisation \\
$\Delta T_{\rm c}$ & increase of the critical temperature \\
$\Delta t$ & integration time step\\
$\delta$ & statistical uncertainty \\
$\epsilon$ & energy parameter \\
$\theta$ & angle of nutation \\
$\xi$ & elementary function \\
$\rho$ & density \\
$\sigma$ & size parameter \\
$\sigma$ & standard deviation \\
$\phi$ & angle of precession \\
$\chi$ & elementary function \\
$\psi$ & elementary function \\
$\omega$ & acentric factor \\
\end{tabular}

\vspace{0.7cm}

\noindent
\textbf{\large Vector properties} \\[0.6cm]
\begin{tabular}{ll}
$\bm r$ & position vector \\
$\bm \omega$ & orientation vector \\ 
\end{tabular}
\clearpage

\noindent
\textbf{\large Subscript} \\[0.6cm]
\begin{tabular}{ll}
c & property at critical point \\
corr & from present correlation \\
EOS & from 2CLJQ hybrid equation of state \\
m & molecular \\
other & from other authors \\
Q & quadrupole \\
sim & from present simulation \\
$\sigma$ & vapour-liquid coexistence \\
0 & reference point \\
2CLJ & two centre Lennard-Jones \\
2CLJQ & two centre Lennard-Jones plus quadrupole \\
\end{tabular}

\vspace{0.7cm}

\noindent
\textbf{\large Superscript} \\[0.6cm]
\begin{tabular}{ll}
*     & reduced \\
$'$   & on bubble line \\
$''$  & on dew line \\
id  & ideal gas \\
res & residual property \\
\end{tabular}

\section{Acknowledgements}
The authors thank Dr. M. Wendland and Dr. M. Mecke, Vienna, for fruitful discussions and Mrs. V. Wetter, Vienna, for her contributions to the simulation work. We gratefully acknowledge financial support by Deutsche Forschungsgemeinschaft, Sonderforschungsbereich 412, University of Stuttgart.

\section{Appendix}

\subsection{Critical properties}

The functions $T^*_{\rm c}(Q^{*2},L^*)$ and $\rho^*_{\rm c}(Q^{*2},L^*)$ were assumed to be linear combinations of elementary functions one of which is a constant $c$, the others depend either on $Q^{*2}$, i.e. $\psi_i \left( Q^{*2} \right)$, or on $L^*$, i.e. $\xi_i \left( L^* \right)$, or on both, i.e. $\chi_i \left( Q^{*2},L^* \right)$. We restricted the number of elementary functions to up to two for both the $Q^{*2}$- and the $L^*$-dependence and to up to four mixed terms. Writing the linear combination of any of the aforementioned functions represented by $y$, we have
\begin{eqnarray}
y \left( Q^{*2},L^* \right) = c + \sum_{i=1}^{\le 2}\alpha_i \cdot \psi_i(Q^{*2}) + \sum_{j=1}^{\le 2}\beta_j \cdot \xi_j(L^*) + \sum_{k=1}^{\le 4} \gamma_k \cdot \chi_k(Q^{*2},L^*).
\label{lincomb}
\end{eqnarray}
The crucial step was the selection of elementary functions $\psi_i$, $\xi_i$, and $\chi_i$. The behaviour of $T^*_{\rm c}$ and $\rho^*_{\rm c}$ for both the $Q^{*2}$- and $L^*$-dependence was taken into account for the selection of the elementary functions for $T^*_{\rm c}(Q^{*2},L^*)$ and $\rho^*_{\rm c}(Q^{*2},L^*)$. The critical data in Table \ref{2CLJQcdomtab} were fitted to these elementary functions in a usual non-weighted least squares sense. The elementary functions and the resulting coefficients are given in Table \ref{corrtab}.

\subsection{Saturated densities, vapour pressure}

The temperature--saturated density correlation is based on Eqs. (\ref{Trho1corrcore}) and (\ref{Trho2corrcore}). The $Q^{*2},L^*$-dependence was ascribed to $T^*_{\rm c}$, $\rho^*_{\rm c}$, and to the coefficients $C_1 \div C''_3$. The correlations for $T^*_{\rm c} \left( Q^{*2},L^* \right)$ and $\rho^*_{\rm c}\left( Q^{*2},L^* \right)$ were used in the optimization procedure for the density temperature correlations $\rho'^*(Q^{*2},L^*,T^*)$ and $\rho''^*(Q^{*2},L^*,T^*)$. The coefficient functions $C_1(Q^{*2},L^*) \div C''_3(Q^{*2},L^*)$ were linear combinations of elementary functions in the sense of Eq. (\ref{lincomb}).
For the correlation $\ln p^*_{\sigma}(Q^{*2},L^*,T^*)$ we restricted the number of coefficient functions to three. Several polynomials in $1/T^*$ and an Antoine-type $\ln p^*_{\sigma}=A+B/(T^*-C)$ were tested. The best results were obtained for the polynomial ansatz
\begin{eqnarray}
\ln p^*_{\sigma}(Q^{*2},L^*,T^*)=c_1(Q^{*2},L^*)+\frac{c_2(Q^{*2},L^*)}{T^*}+\frac{c_3(Q^{*2},L^*)}{T^{*4}}, \nonumber
\label{pTcorr}
\end{eqnarray}
which is in accordance to former investigations of Lotfi et al. \cite{lotfi9213} and Kriebel et al. \cite{kriebel9538}. Again, the coefficients $c_1(Q^{*2},L^*) \div c_3(Q^{*2},L^*)$ were assumed to be linear combinations of the elementary functions in the sense of Eq. (\ref{lincomb}).

In principle, in developping the correlations for the saturated densities and the vapour pressure, we filtered out elementary functions in the same way as we have done it for the correlation of critical data. Fits of VLE data from simulation and from the 2CLJQ EOS for some model fluids provided useful information about the topology of the coefficients in $Q^{*2}$- and $L^*$-direction. When an initial set of elementary functions is found, the coefficients of these elementary functions can be determined by weighted least squares minimization of $F$ and $G$ defined as
\begin{eqnarray}
F &=& \sum_i \left[ \phantom{+} \frac{1}{ \delta \rho'^{*2}_i} \left[ \rho'^* \left( Q^{*2}_i,L^*_i,T^*_i \right) - \rho'^*_{{\rm sim},i} \right]^2 \right. \nonumber \\ \nonumber
&\phantom{=}& \phantom{\sum_i} + \left. \frac{1}{ \delta \rho''^{*2}_i} \left[ \rho''^* \left( Q^{*2}_i,L^*_i,T^*_i \right) - \rho''^*_{{\rm sim},i} \right]^2 \right] \stackrel{\rm !}{=} {\rm min}, \\ \nonumber
G &=& \sum_i \frac{1}{ \left( \delta {\rm ln}p_{\sigma,i} \right)^2} \left[ {\rm ln}p_{\sigma} \left( Q^{*2}_i,L^*_i,T^*_i \right) -{\rm ln}p_{\sigma,{\rm sim},i} \right]^2 \stackrel{\rm !}{=} {\rm min}. \nonumber
\end{eqnarray}
$F$ and $G$ were also used to optimize the initial set of elementary functions. The resulting elementary functions and their coefficients are given in Table \ref{corrtab}.


\clearpage

\listoftables
\clearpage

\begin{table}[ht]
\noindent
\caption[Vapour-liquid equilibrium data. Extract from simulation results for 30 model fluids for low, mid, and high temperatures. At higher temperatures, marked by \dag, data are based completely on simulations. For the remaining temperatures the second virial coefficient was used for the vapour phase. The numbers in parentheses are the uncertainties of the last decimal digits indicated.]{}
\label{t2cljqvle}
\medskip
\begin{center}
{\footnotesize
\begin{tabular}{l|ccccc}\hline\hline
{\normalsize ~$\!T^*$} &
{\normalsize $p_{\sigma}^*$} &
{\normalsize $\rho'^*$} & {\normalsize $\rho''^*$} &
{\normalsize $h'^{\mbox{\scriptsize res*}}$} &
{\normalsize $h''^{\mbox{\scriptsize res*}}$} \\ \hline
\multicolumn{6}{l}{{\normalsize $L^*=0$, $Q^{*2}=0$}} \\ \hline
2.92600 &   0.00852    (33) &   0.82938    (25) &   0.00299    (12) & -26.8247\0   (86) &  -0.2197\0   (89) \\
4.25600\dag &   0.1469\0   (14) &   0.66404    (55) &   0.04347    (53) & -22.457\0\0  (15) &  -2.459\0\0  (33) \\
5.05400\dag &   0.4120\0   (34) &   0.4925\0   (31) &   0.1475\0   (38) & -17.768\0\0  (77) &  -6.98\0\0\0 (17) \\ \hline
\multicolumn{6}{l}{{\normalsize $L^*=0$, $Q^{*2}=1$}} \\ \hline
2.97000 &   0.00796    (49) &   0.84416    (22) &   0.00275    (18) & -28.4469\0   (83) &  -0.212\0\0  (14) \\
4.32000\dag &   0.1472\0   (13) &   0.67461    (58) &   0.04239    (62) & -23.448\0\0  (17) &  -2.453\0\0  (43) \\
5.13000\dag &   0.4108\0   (34) &   0.4994\0   (24) &   0.1398\0   (56) & -18.453\0\0  (60) &  -6.84\0\0\0 (26) \\ \hline
\multicolumn{6}{l}{{\normalsize $L^*=0$, $Q^{*2}=2$}} \\ \hline
3.09100 &   0.0057\0   (11) &   0.87669    (24) &   0.00186    (36) & -32.4004\0   (97) &  -0.165\0\0  (31) \\
4.49600\dag &   0.1417\0   (17) &   0.69986    (61) &   0.03877    (86) & -26.051\0\0  (19) &  -2.490\0\0  (77) \\
5.33900\dag &   0.4199\0   (36) &   0.5288\0   (22) &   0.1375\0   (31) & -20.553\0\0  (59) &  -7.27\0\0\0 (16) \\ \hline
\multicolumn{6}{l}{{\normalsize $L^*=0$, $Q^{*2}=3$}} \\ \hline
3.26700 &   0.0046\0   (11) &   0.91399    (29) &   0.00144    (35) & -37.714\0\0  (14) &  -0.155\0\0  (37) \\
4.75200\dag &   0.1352\0   (16) &   0.73335    (52) &   0.03653    (67) & -29.885\0\0  (19) &  -2.745\0\0  (76) \\
5.64300\dag &   0.4213\0   (32) &   0.5667\0   (17) &   0.1220\0   (20) & -23.694\0\0  (53) &  -7.09\0\0\0 (11) \\ \hline
\multicolumn{6}{l}{{\normalsize $L^*=0$, $Q^{*2}=4$}} \\ \hline
3.47600 &   0.0036\0   (14) &   0.95228    (25) &   0.00106    (43) & -43.992\0\0  (13) &  -0.144\0\0  (58) \\
5.05600\dag &   0.1202\0   (23) &   0.77053    (46) &   0.0260\0   (19) & -34.733\0\0  (19) &  -2.04\0\0\0 (27) \\
6.00400\dag &   0.4186\0   (42) &   0.6079\0   (15) &   0.1082\0   (27) & -27.730\0\0  (54) &  -7.09\0\0\0 (18) \\ \hline
\multicolumn{6}{l}{{\normalsize $L^*=0.2$, $Q^{*2}=0$}} \\ \hline
2.42000 &   0.00576    (31) &   0.73095    (24) &   0.00244    (13) & -22.5710\0   (79) &  -0.1737\0   (96) \\
3.52000\dag &   0.1092\0   (11) &   0.58419    (52) &   0.03850    (56) & -18.792\0\0  (14) &  -2.024\0\0  (47) \\
4.18000\dag &   0.2990\0   (22) &   0.4236\0   (23) &   0.1279\0   (23) & -14.595\0\0  (52) &  -5.769\0\0  (94) \\ \hline
\multicolumn{6}{l}{{\normalsize $L^*=0.2$, $Q^{*2}=1$}} \\ \hline
2.45850 &   0.00514    (39) &   0.74117    (23) &   0.00214    (17) & -23.7142\0   (79) &  -0.158\0\0  (12) \\
3.57600\dag &   0.1092\0   (12) &   0.58928    (50) &   0.03927    (80) & -19.412\0\0  (14) &  -2.219\0\0  (69) \\
4.24650\dag &   0.3123\0   (23) &   0.4320\0   (30) &   0.1317\0   (26) & -15.102\0\0  (69) &  -6.04\0\0\0 (13) \\ \hline
\multicolumn{6}{l}{{\normalsize $L^*=0.2$, $Q^{*2}=2$}} \\ \hline
2.55200 &   0.00458    (49) &   0.76427    (20) &   0.00183    (20) & -26.5697\0   (78) &  -0.151\0\0  (17) \\
3.71200\dag &   0.1062\0   (12) &   0.60533    (59) &   0.03580    (38) & -21.203\0\0  (17) &  -2.147\0\0  (24) \\
4.40800\dag &   0.3213\0   (27) &   0.4448\0   (22) &   0.1298\0   (27) & -16.348\0\0  (56) &  -6.21\0\0\0 (11) \\ \hline
\multicolumn{6}{l}{{\normalsize $L^*=0.2$, $Q^{*2}=3$}} \\ \hline
2.68950 &   0.00300    (60) &   0.79142    (26) &   0.00113    (23) & -30.474\0\0  (11) &  -0.110\0\0  (22) \\
3.91200\dag &   0.1081\0   (16) &   0.62931    (54) &   0.03498    (66) & -23.918\0\0  (18) &  -2.353\0\0  (57) \\
4.64550\dag &   0.3234\0   (27) &   0.4559\0   (26) &   0.1147\0   (41) & -18.074\0\0  (70) &  -6.11\0\0\0 (21) \\ \hline
\end{tabular}}
\end{center}
\end{table}

\begin{table}[ht]
\noindent
\begin{center}
Table \ref{t2cljqvle}: continued. \\
\end{center}\medskip
\begin{center}
{\footnotesize
\begin{tabular}{l|ccccc}\hline\hline
{\normalsize ~$\!T^*$} &
{\normalsize $p_{\sigma}^*$} &
{\normalsize $\rho'^*$} & {\normalsize $\rho''^*$} &
{\normalsize $h'^{\mbox{\scriptsize res*}}$} &
{\normalsize $h''^{\mbox{\scriptsize res*}}$} \\ \hline
\multicolumn{6}{l}{{\normalsize $L^*=0.2$, $Q^{*2}=4$}} \\ \hline
2.86000 &   0.0044\0   (15) &   0.81978    (22) &   0.00158    (53) & -35.098\0\0  (11) &  -0.187\0\0  (63) \\
4.16000\dag &   0.1061\0   (17) &   0.65261    (48) &   0.0281\0   (14) & -27.201\0\0  (18) &  -1.94\0\0\0 (15) \\
4.94000\dag &   0.3428\0   (31) &   0.4877\0   (20) &   0.1137\0   (42) & -20.837\0\0  (61) &  -6.60\0\0\0 (24) \\ \hline
\multicolumn{6}{l}{{\normalsize $L^*=0.4$, $Q^{*2}=0$}} \\ \hline
1.78750 &   0.00325    (21) &   0.59531    (20) &   0.00187    (13) & -17.2882\0   (61) &  -0.1312\0   (89) \\
2.60000\dag &   0.06524    (64) &   0.47244    (58) &   0.03277    (80) & -14.213\0\0  (14) &  -1.708\0\0  (73) \\
3.08750\dag &   0.1892\0   (18) &   0.3417\0   (26) &   0.1231\0   (59) & -10.955\0\0  (54) &  -5.08\0\0\0 (21) \\ \hline
\multicolumn{6}{l}{{\normalsize $L^*=0.4$, $Q^{*2}=1$}} \\ \hline
1.81500 &   0.00267    (19) &   0.60158    (20) &   0.00150    (11) & -18.0097\0   (64) &  -0.1077\0   (76) \\
2.64000\dag &   0.06547    (71) &   0.47589    (47) &   0.03121    (59) & -14.605\0\0  (12) &  -1.626\0\0  (41) \\
3.13500\dag &   0.1910\0   (20) &   0.3273\0   (46) &   0.1163\0   (41) & -10.845\0\0  (96) &  -4.90\0\0\0 (16) \\ \hline
\multicolumn{6}{l}{{\normalsize $L^*=0.4$, $Q^{*2}=2$}} \\ \hline
1.88100 &   0.00226    (33) &   0.61695    (19) &   0.00122    (18) & -19.8956\0   (69) &  -0.096\0\0  (14) \\
2.73600\dag &   0.06747    (71) &   0.48583    (52) &   0.03174    (31) & -15.731\0\0  (14) &  -1.772\0\0  (22) \\
3.24900\dag &   0.2002\0   (20) &   0.3332\0   (39) &   0.1166\0   (39) & -11.498\0\0  (91) &  -5.12\0\0\0 (16) \\ \hline
\multicolumn{6}{l}{{\normalsize $L^*=0.4$, $Q^{*2}=3$}} \\ \hline
1.97450 &   0.00308    (45) &   0.63581    (17) &   0.00160    (24) & -22.5113\0   (73) &  -0.144\0\0  (22) \\
2.87200\dag &   0.06607    (81) &   0.49850    (43) &   0.02904    (41) & -17.378\0\0  (13) &  -1.759\0\0  (33) \\
3.41050\dag &   0.2118\0   (21) &   0.3515\0   (25) &   0.1200\0   (36) & -12.817\0\0  (63) &  -5.69\0\0\0 (16) \\ \hline
\multicolumn{6}{l}{{\normalsize $L^*=0.4$, $Q^{*2}=4$}} \\ \hline
2.09000 &   0.0023\0   (11) &   0.65625    (20) &   0.00112    (54) & -25.654\0\0  (10) &  -0.118\0\0  (57) \\
3.04000\dag &   0.0670\0   (11) &   0.51516    (44) &   0.02688    (71) & -19.512\0\0  (15) &  -1.773\0\0  (58) \\
3.61000\dag &   0.2246\0   (21) &   0.3635\0   (20) &   0.1184\0   (34) & -14.256\0\0  (51) &  -6.07\0\0\0 (16) \\ \hline
\multicolumn{6}{l}{{\normalsize $L^*=0.505$, $Q^{*2}=0$}} \\ \hline
1.55650 &   0.00275    (15) &   0.54062    (17) &   0.00182    (10) & -15.3895\0   (52) &  -0.1289\0   (72) \\
2.26400\dag &   0.05086    (49) &   0.42814    (44) &   0.02815    (29) & -12.583\0\0  (11) &  -1.385\0\0  (17) \\
2.68850\dag &   0.1487\0   (15) &   0.2927\0   (50) &   0.1093\0   (38) &  -9.31\0\0\0 (10) &  -4.42\0\0\0 (14) \\ \hline
\multicolumn{6}{l}{{\normalsize $L^*=0.505$, $Q^{*2}=1$}} \\ \hline
1.57850 &   0.00219    (20) &   0.54667    (17) &   0.00142    (14) & -16.0333\0   (55) &  -0.1021\0   (97) \\
2.29600\dag &   0.05150    (62) &   0.43263    (51) &   0.02783    (28) & -12.959\0\0  (13) &  -1.381\0\0  (13) \\
2.72650\dag &   0.1525\0   (16) &   0.2972\0   (64) &   0.1123\0   (40) &  -9.55\0\0\0 (12) &  -4.59\0\0\0 (15) \\ \hline
\multicolumn{6}{l}{{\normalsize $L^*=0.505$, $Q^{*2}=2$}} \\ \hline
1.63350 &   0.00197    (51) &   0.56070    (18) &   0.00123    (33) & -17.6851\0   (69) &  -0.097\0\0  (26) \\
2.37600\dag &   0.05095    (61) &   0.44022    (43) &   0.02713    (70) & -13.902\0\0  (12) &  -1.470\0\0  (50) \\
2.82150\dag &   0.1611\0   (19) &   0.3072\0   (31) &   0.1142\0   (42) & -10.273\0\0  (72) &  -4.83\0\0\0 (15) \\ \hline
\end{tabular}}
\end{center}
\end{table}

\begin{table}[ht]
\noindent
\begin{center}
Table \ref{t2cljqvle}: continued. \\
\end{center}
\medskip
\begin{center}
{\footnotesize
\begin{tabular}{l|ccccc}\hline\hline
{\normalsize ~$\!T^*$} &
{\normalsize $p_{\sigma}^*$} &
{\normalsize $\rho'^*$} & {\normalsize $\rho''^*$} &
{\normalsize $h'^{\mbox{\scriptsize res*}}$} &
{\normalsize $h''^{\mbox{\scriptsize res*}}$} \\ \hline
\multicolumn{6}{l}{{\normalsize $L^*=0.505$, $Q^{*2}=3$}} \\ \hline
1.71600 &   0.00153    (37) &   0.57667    (18) &   0.00091    (22) & -19.9069\0   (70) &  -0.081\0\0  (20) \\
2.49600\dag &   0.05254    (69) &   0.45184    (47) &   0.0277\0   (10) & -15.321\0\0  (14) &  -1.686\0\0  (96) \\
2.96400\dag &   0.1684\0   (21) &   0.3064\0   (36) &   0.1138\0   (54) & -10.964\0\0  (90) &  -5.17\0\0\0 (22) \\ \hline
\multicolumn{6}{l}{{\normalsize $L^*=0.505$, $Q^{*2}=4$}} \\ \hline
1.81500 &   0.00089    (50) &   0.59445    (18) &   0.00049    (28) & -22.5908\0   (89) &  -0.052\0\0  (30) \\
2.64000\dag &   0.05430    (87) &   0.46561    (47) &   0.02598    (77) & -17.109\0\0  (16) &  -1.672\0\0  (69) \\
3.13500\dag &   0.1781\0   (19) &   0.3220\0   (31) &   0.1099\0   (37) & -12.274\0\0  (84) &  -5.45\0\0\0 (18) \\ \hline
\multicolumn{6}{l}{{\normalsize $L^*=0.6$, $Q^{*2}=0$}} \\ \hline
1.40250 &   0.00202    (15) &   0.50020    (15) &   0.00148    (12) & -14.0888\0   (46) &  -0.1052\0   (82) \\
2.04000\dag &   0.04288    (48) &   0.39339    (41) &   0.02746    (70) & -11.408\0\0  (10) &  -1.387\0\0  (55) \\
2.42250\dag &   0.1286\0   (13) &   0.2643\0   (34) &   0.1075\0   (33) &  -8.299\0\0  (70) &  -4.26\0\0\0 (13) \\ \hline
\multicolumn{6}{l}{{\normalsize $L^*=0.6$, $Q^{*2}=1$}} \\ \hline
1.41900 &   0.00173    (16) &   0.50637    (16) &   0.00125    (11) & -14.7048\0   (52) &  -0.0909\0   (83) \\
2.06400\dag &   0.04212    (47) &   0.39815    (47) &   0.02676    (68) & -11.785\0\0  (12) &  -1.407\0\0  (55) \\
2.45100\dag &   0.1301\0   (14) &   0.2800\0   (20) &   0.1153\0   (58) &  -8.804\0\0  (41) &  -4.55\0\0\0 (18) \\ \hline
\multicolumn{6}{l}{{\normalsize $L^*=0.6$, $Q^{*2}=2$}} \\ \hline
1.46850 &   0.00136    (23) &   0.51998    (17) &   0.00094    (16) & -16.2362\0   (63) &  -0.075\0\0  (13) \\
2.13600\dag &   0.04282    (62) &   0.40725    (46) &   0.0307\0   (19) & -12.696\0\0  (13) &  -1.89\0\0\0 (17) \\
2.53650\dag &   0.1311\0   (12) &   0.2598\0   (65) &   0.1016\0   (32) &  -8.79\0\0\0 (14) &  -4.32\0\0\0 (12) \\ \hline
\multicolumn{6}{l}{{\normalsize $L^*=0.6$, $Q^{*2}=3$}} \\ \hline
1.54000 &   0.00144    (25) &   0.53541    (17) &   0.00095    (17) & -18.2738\0   (74) &  -0.086\0\0  (15) \\
2.24000\dag &   0.04270    (57) &   0.41820    (41) &   0.02300    (33) & -13.986\0\0  (13) &  -1.289\0\0  (25) \\
2.66000\dag &   0.1398\0   (15) &   0.2792\0   (44) &   0.1058\0   (48) &  -9.87\0\0\0 (10) &  -4.73\0\0\0 (19) \\ \hline
\multicolumn{6}{l}{{\normalsize $L^*=0.6$, $Q^{*2}=4$}} \\ \hline
1.62800 &   0.00083    (32) &   0.55127    (21) &   0.00052    (20) & -20.655\0\0  (11) &  -0.055\0\0  (22) \\
2.36800\dag &   0.04378    (81) &   0.43014    (48) &   0.02306    (70) & -15.565\0\0  (16) &  -1.495\0\0  (68) \\
2.81200\dag &   0.1487\0   (16) &   0.2959\0   (20) &   0.1028\0   (31) & -11.108\0\0  (55) &  -4.99\0\0\0 (14) \\ \hline
\multicolumn{6}{l}{{\normalsize $L^*=0.8$, $Q^{*2}=0$}} \\ \hline
1.17700 &   0.00143    (15) &   0.43844    (13) &   0.00124    (14) & -12.2840\0   (39) &  -0.090\0\0  (10) \\
1.71200\dag &   0.03152    (36) &   0.34003    (50) &   0.02282    (22) &  -9.768\0\0  (12) &  -1.090\0\0  (12) \\
2.03300\dag &   0.0972\0   (11) &   0.2103\0   (44) &   0.0969\0   (32) &  -6.664\0\0  (91) &  -3.71\0\0\0 (11) \\ \hline
\multicolumn{6}{l}{{\normalsize $L^*=0.8$, $Q^{*2}=1$}} \\ \hline
1.18800 &   0.00120    (18) &   0.44763    (13) &   0.00103    (16) & -13.0246\0   (43) &  -0.078\0\0  (12) \\
1.72800\dag &   0.02961    (36) &   0.34908    (40) &   0.02164    (26) & -10.278\0\0  (10) &  -1.099\0\0  (12) \\
2.05200\dag &   0.0922\0   (10) &   0.2460\0   (15) &   0.0874\0   (24) &  -7.628\0\0  (38) &  -3.493\0\0  (83) \\ \hline
\end{tabular}}
\end{center}
\end{table}
\clearpage

\begin{table}[ht]
\noindent
\begin{center}
Table \ref{t2cljqvle}: continued. \\
\end{center}
\medskip
\begin{center}
{\footnotesize
\begin{tabular}{l|ccccc}\hline\hline
{\normalsize ~$\!T^*$} &
{\normalsize $p_{\sigma}^*$} &
{\normalsize $\rho'^*$} & {\normalsize $\rho''^*$} &
{\normalsize $h'^{\mbox{\scriptsize res*}}$} &
{\normalsize $h''^{\mbox{\scriptsize res*}}$} \\ \hline
\multicolumn{6}{l}{{\normalsize $L^*=0.8$, $Q^{*2}=2$}} \\ \hline
1.22650 &   0.00083    (20) &   0.46255    (15) &   0.00069    (17) & -14.5252\0   (60) &  -0.058\0\0  (15) \\
1.78400\dag &   0.02859    (42) &   0.36096    (43) &   0.01992    (37) & -11.232\0\0  (12) &  -1.109\0\0  (19) \\
2.11850\dag &   0.09230    (95) &   0.2449\0   (25) &   0.0813\0   (22) &  -8.067\0\0  (56) &  -3.461\0\0  (87) \\ \hline
\multicolumn{6}{l}{{\normalsize $L^*=0.8$, $Q^{*2}=3$}} \\ \hline
1.27600 &   0.00053    (16) &   0.47977    (14) &   0.00042    (13) & -16.4793\0   (60) &  -0.043\0\0  (13) \\
1.85600\dag &   0.02629    (53) &   0.37721    (36) &   0.0144\0   (11) & -12.622\0\0  (12) &  -0.76\0\0\0 (11) \\
2.20400\dag &   0.0887\0   (10) &   0.2694\0   (15) &   0.0699\0   (34) &  -9.309\0\0  (38) &  -3.29\0\0\0 (17) \\ \hline
\multicolumn{6}{l}{{\normalsize $L^*=0.8$, $Q^{*2}=4$}} \\ \hline
1.34200 &   0.00033    (26) &   0.49548    (17) &   0.00025    (19) & -18.6173\0   (83) &  -0.030\0\0  (24) \\
1.95200\dag &   0.02440    (67) &   0.39236    (38) &   0.01619    (54) & -14.226\0\0  (13) &  -1.157\0\0  (71) \\
2.31800\dag &   0.09208    (95) &   0.2893\0   (14) &   0.0684\0   (15) & -10.673\0\0  (36) &  -3.521\0\0  (74) \\ \hline
\end{tabular}}
\end{center}
\end{table}
\clearpage
\begin{table}[ht]
\noindent
\caption[Critical data, the compressibility factor, and the acentric factor of 30 2CLJQ model fluids.]{}
\label{2CLJQcdomtab}
\bigskip
\begin{center}
\begin{tabular}{|c|c||l|l|l|l|l|l||l|}\hline
\multicolumn{2}{|c||}{ } & \multicolumn{6}{|l||}{$L^*$} & \\ \cline{3-8}
\multicolumn{2}{|c||}{ } & $0$ & $0.2$ & $0.4$ & $0.505$ & $0.6$ & $0.8$ & \\ \hline \hline
$Q^{*2}$ & 0 & \phantom{-}5.236  & \phantom{-}4.313  & \phantom{-}3.163  & \phantom{-}2.735  & \phantom{-}2.454  & \phantom{-}2.049  & $T^*_{\rm c}$ \\ \cline{3-9}
         &   & \phantom{-}0.3143 & \phantom{-}0.2740 & \phantom{-}0.2251 & \phantom{-}0.2032 & \phantom{-}0.1850 & \phantom{-}0.1577 & $\rho^*_{\rm c}$ \\ \cline{3-9}
         &   & \phantom{-}0.4757 & \phantom{-}0.3573 & \phantom{-}0.2103 & \phantom{-}0.1680 & \phantom{-}0.1377 & \phantom{-}0.1006 & $p^*_{\rm c}$ \\ \cline{3-9}
         &   & \phantom{-}0.2891 & \phantom{-}0.3023 & \phantom{-}0.2954 & \phantom{-}0.3023 & \phantom{-}0.3033 & \phantom{-}0.3113 & $Z_{\rm c}$ \\ \cline{3-9}
         &   &           -0.0551 &           -0.0194 & \phantom{-}0.0319 & \phantom{-}0.0439 & \phantom{-}0.0553 & \phantom{-}0.0731 & $\omega$ \\ \cline{2-9}
         & 1 & \phantom{-}5.312  & \phantom{-}4.388  & \phantom{-}3.195  & \phantom{-}2.771  & \phantom{-}2.499  & \phantom{-}2.097  & $T^*_{\rm c}$ \\ \cline{3-9}
         &   & \phantom{-}0.3169 & \phantom{-}0.2796 & \phantom{-}0.2259 & \phantom{-}0.2036 & \phantom{-}0.1863 & \phantom{-}0.1619 & $\rho^*_{\rm c}$ \\ \cline{3-9}
         &   & \phantom{-}0.4876 & \phantom{-}0.3664 & \phantom{-}0.2133 & \phantom{-}0.1701 & \phantom{-}0.1394 & \phantom{-}0.1001 & $p^*_{\rm c}$ \\ \cline{3-9}
         &   & \phantom{-}0.2897 & \phantom{-}0.2986 & \phantom{-}0.2955 & \phantom{-}0.3015 & \phantom{-}0.2994 & \phantom{-}0.2948 & $Z_{\rm c}$ \\ \cline{3-9}
         &   &           -0.0354 &           -0.0029 & \phantom{-}0.0470 & \phantom{-}0.0598 & \phantom{-}0.0733 & \phantom{-}0.1015 & $\omega$ \\ \cline{2-9}
         & 2 & \phantom{-}5.575  & \phantom{-}4.554  & \phantom{-}3.316  & \phantom{-}2.875  & \phantom{-}2.580  & \phantom{-}2.160  & $T^*_{\rm c}$\\ \cline{3-9}
         &   & \phantom{-}0.3197 & \phantom{-}0.2832 & \phantom{-}0.2256 & \phantom{-}0.2090 & \phantom{-}0.1888 & \phantom{-}0.1652 & $\rho^*_{\rm c}$ \\ \cline{3-9}
         &   & \phantom{-}0.5187 & \phantom{-}0.3899 & \phantom{-}0.2221 & \phantom{-}0.1768 & \phantom{-}0.1452 & \phantom{-}0.1019 & $p^*_{\rm c}$ \\ \cline{3-9}
         &   & \phantom{-}0.2910 & \phantom{-}0.3023 & \phantom{-}0.2969 & \phantom{-}0.2942 & \phantom{-}0.2981 & \phantom{-}0.2856 & $Z_{\rm c}$ \\ \cline{3-9}
         &   & \phantom{-}0.0104 & \phantom{-}0.0373 & \phantom{-}0.0844 & \phantom{-}0.0974 & \phantom{-}0.1133 & \phantom{-}0.1571 & $\omega$ \\ \cline{2-9}
         & 3 & \phantom{-}5.968  & \phantom{-}4.807  & \phantom{-}3.489  & \phantom{-}3.007  & \phantom{-}2.695  & \phantom{-}2.266  & $T^*_{\rm c}$ \\ \cline{3-9}
         &   & \phantom{-}0.3303 & \phantom{-}0.2880 & \phantom{-}0.2309 & \phantom{-}0.2097 & \phantom{-}0.1940 & \phantom{-}0.1680 & $\rho^*_{\rm c}$ \\ \cline{3-9}
         &   & \phantom{-}0.5617 & \phantom{-}0.4221 & \phantom{-}0.2362 & \phantom{-}0.1880 & \phantom{-}0.1550 & \phantom{-}0.1079 & $p^*_{\rm c}$ \\ \cline{3-9}
         &   & \phantom{-}0.2849 & \phantom{-}0.3049 & \phantom{-}0.2932 & \phantom{-}0.2981 & \phantom{-}0.2965 & \phantom{-}0.2834 & $Z_{\rm c}$ \\ \cline{3-9}
         &   & \phantom{-}0.0628 & \phantom{-}0.0877 & \phantom{-}0.1343 & \phantom{-}0.1464 & \phantom{-}0.1620 & \phantom{-}0.2129 & $\omega$ \\ \cline{2-9}
         & 4 & \phantom{-}6.398  & \phantom{-}5.143  & \phantom{-}3.692  & \phantom{-}3.195  & \phantom{-}2.866  & \phantom{-}2.408  & $T^*_{\rm c}$ \\ \cline{3-9}
         &   & \phantom{-}0.3395 & \phantom{-}0.2908 & \phantom{-}0.2358 & \phantom{-}0.2143 & \phantom{-}0.1960 & \phantom{-}0.1710 & $\rho^*_{\rm c}$ \\ \cline{3-9}
         &   & \phantom{-}0.5933 & \phantom{-}0.4499 & \phantom{-}0.2524 & \phantom{-}0.2005 & \phantom{-}0.1646 & \phantom{-}0.1135 & $p^*_{\rm c}$ \\ \cline{3-9}
         &   & \phantom{-}0.2731 & \phantom{-}0.3008 & \phantom{-}0.2899 & \phantom{-}0.2928 & \phantom{-}0.2930 & \phantom{-}0.2756 & $Z_{\rm c}$ \\ \cline{3-9}
         &   & \phantom{-}0.1070 & \phantom{-}0.1378 & \phantom{-}0.1914 & \phantom{-}0.2032 & \phantom{-}0.2174 & \phantom{-}0.2649 & $\omega$ \\ \hline
\end{tabular}
\end{center}
\end{table}
\clearpage
\begin{table}[ht]
\noindent
\caption[Elementary functions and their coefficients for the correlations $T^*_{\rm c} \left( Q^{*2},L^* \right)$, $\rho^*_{\rm c} \left( Q^{*2},L^* \right)$, $C_1\left( Q^{*2},L^* \right)$ to $C''_3\left( Q^{*2},L^* \right)$, and $c_1\left( Q^{*2},L^* \right)$ to $c_3\left( Q^{*2},L^* \right)$. The notation is simplified: $\ell$ is $L^*$, $q$ is $Q^{*2}$.]{}
\label{corrtab}
\bigskip
\begin{tabular}{|l|l|l|l|l|l|}\hline
\multicolumn{3}{|l|}{$T^*_{\rm c}\left(q,\ell \right)$} & \multicolumn{3}{|l|}{$\rho^*_{\rm c}\left(q,\ell \right)$} \\ \hline
$c$      & 1 & $\phantom{+}0.15075788232 \cdot 10^{1}$ & $c$ & 1 & $\phantom{+}0.31431713778$ \\ \hline
$\psi_i$ & $q^2$ & $\phantom{+}0.20472313140 \cdot 10^{-1}$ & $\psi_i$ & $q^2$ & $\phantom{+}0.24699993812 \cdot 10^{-2}$ \\ \cline{2-3} \cline{5-6}
         & $q^3$ & $-0.12916709336 \cdot 10^{-2}$ & & $q^3$ & $-0.24220105645 \cdot 10^{-3}$ \\ \cline{1-3} \cline{4-6}
$\xi_i$  & $1/(0.1+\ell^2)$ & $\phantom{+}0.33194556111$ & $\xi_i$ & $\ell^2/(0.11+\ell^2)$ & $-0.14520352142$ \\ \cline{2-3} \cline{5-6}
         & $1/(0.1+\ell^5)$ & $\phantom{+}0.41364616282 \cdot 10^{-1}$ & & $\ell^5/(0.11+\ell^5)$ & $-0.42590980830 \cdot 10^{-1}$ \\ \cline{1-3} \cline{4-6}
$\chi_i$ & $q^2/(0.1+\ell^2)$ & $\phantom{+}0.97556493849 \cdot 10^{-2}$ & $\chi_i$ & $\ell^2 q^2/(0.11+\ell^2)$ & $-0.27008826493 \cdot 10^{-2}$\\ \cline{2-3} \cline{5-6}
         & $q^2/(0.1+\ell^5)$ & $-0.17158397457 \cdot 10^{-2}$ & & $\ell^5 q^2/(0.11+\ell^5)$ & $\phantom{+}0.27854853704 \cdot 10^{-2}$ \\ \cline{2-3} \cline{5-6}
         & $q^3/(0.1+\ell^2)$ & $-0.81735779939 \cdot 10^{-3}$ & & $\ell^2 q^3/(0.11+\ell^2)$ & $\phantom{+}0.30075661416 \cdot 10^{-3}$ \\ \cline{2-3} \cline{5-6}
         & $q^3/(0.1+\ell^5)$ & $\phantom{+}0.23012291309 \cdot 10^{-3}$ & & $\ell^5 q^3/(0.11+\ell^5)$ & $-0.50847561388 \cdot 10^{-3}$ \\ \cline{1-3} \cline{4-6}
\multicolumn{3}{|l|}{$C_1\left(q,\ell \right)$} & \multicolumn{3}{|l|}{$C'_2\left(q,\ell \right)$} \\ \hline
$c$      & 1 & $\phantom{+}0.30195487445$ & $c$ & 1 & $\phantom{+}0.49559557035 \cdot 10^{-1}$ \\ \hline
$\psi_i$ & $q^2$ & $-0.37967456269 \cdot 10^{-3}$ & $\psi_i$ & $q^2$ & $\phantom{+}0.61257929294 \cdot 10^{-2}$ \\ \cline{2-3} \cline{5-6}
         & $q^3$ & $\phantom{+}0.67459200881 \cdot 10^{-3}$ & & $q^3$ & $-0.16842248337 \cdot 10^{-2}$ \\ \cline{1-3} \cline{4-6}
$\xi_i$  & $\ell^3/(\ell+0.4)^3$ & $-0.16082575198 \cdot 10^{-1}$ & $\xi_i$ & $\ell^2$ & $\phantom{+}0.76651784089 \cdot 10^{-1}$ \\ \cline{2-3} \cline{5-6}
         & $\ell^4/(\ell+0.4)^5$ & $-0.52279636608$ & & $\ell^3$ & $\phantom{+}0.43209147006 \cdot 10^{-1}$ \\ \cline{1-3} \cline{4-6}
$\chi_i$ & $q^2 \ell^2/(\ell+0.4)$ & $\phantom{+}0.11220680075 \cdot 10^{-1}$ & $\chi_i$ & $q^2 \ell^2$ & $-0.14895101590 \cdot 10^{-1}$\\ \cline{2-3} \cline{5-6}
         & $q^2 \ell^3/(\ell+0.4)^7$ & $-0.45237300196 \cdot 10^{-2}$ & & $q^2 \ell^3$ & $-0.21199303602 \cdot 10^{-2}$ \\ \cline{2-3} \cline{5-6}
         & $q^3 \ell^2/(\ell+0.4)$ & $-0.27116058021 \cdot 10^{-2}$ & & $q^3 \ell^2$ & $\phantom{+}0.30307914217 \cdot 10^{-2}$ \\ \cline{2-3} \cline{5-6}
         & $q^3 \ell^3/(\ell+0.4)^7$ & $\phantom{+}0.39497438901 \cdot 10^{-3}$ & & & \\ \cline{1-3} \cline{4-6}
\multicolumn{3}{|l|}{$C'_3\left(q,\ell \right)$} & \multicolumn{3}{|l|}{$C''_2\left(q,\ell \right)$} \\ \hline
$c$      & 1 & $\phantom{+}0.66767184355 \cdot 10^{-3}$ & $c$ & 1 & $\phantom{+}0.19846809576 \cdot 10^{-1}$ \\ \hline
$\psi_i$ & $q^2$ & $-0.21255206238 \cdot 10^{-2}$ & $\psi_i$ & $q^2$ & $-0.36770421816 \cdot 10^{-2}$ \\ \cline{2-3} \cline{5-6}
         & $q^3$ & $\phantom{+}0.55917439122 \cdot 10^{-3}$ & & $q^3$ & $\phantom{+}0.12514990954 \cdot 10^{-2}$ \\ \cline{1-3} \cline{4-6}
$\xi_i$  & $\ell$ & $-0.50078939740 \cdot 10^{-2}$ & $\xi_i$ & $\ell^2$ & $-0.59416808329 \cdot 10^{-1}$ \\ \cline{2-3} \cline{5-6}
         & $\ell^4$ & $-0.70891584543 \cdot 10^{-1}$ & & $\ell^3$ & $\phantom{+}0.52359791216 \cdot 10^{-1}$ \\ \cline{1-3} \cline{4-6}
$\chi_i$ & $q^2 \ell$ & $\phantom{+}0.12188553188 \cdot 10^{-2}$ & $\chi_i$ & $q^2 \ell^2$ & $\phantom{+}0.19662720492 \cdot 10^{-1}$\\ \cline{2-3} \cline{5-6}
         & $q^2 \ell^4$ & $\phantom{+}0.86157078089 \cdot 10^{-2}$ & & $q^2 \ell^3$ & $-0.77114540331 \cdot 10^{-2}$ \\ \cline{2-3} \cline{5-6}
         & $q^3 \ell^4$ & $-0.14075623914 \cdot 10^{-2}$ & & $q^3 \ell^2$ & $-0.26044833098 \cdot 10^{-2}$ \\ \hline
\end{tabular}
\end{table}

\begin{table}[ht]
\begin{center}
Table \ref{corrtab}: continued. \\
\end{center}
\medskip
\begin{tabular}{|l|l|l|l|l|l|}\hline
\multicolumn{3}{|l|}{$C''_3\left(q,\ell \right)$} & \multicolumn{3}{|l|}{$c_1\left(q,\ell \right)$} \\ \hline
$c$      & 1 & $\phantom{+}0.11857649302 \cdot 10^{-1}$ & $c$ & 1 & $\phantom{+}0.43338821199 \cdot 10^1$ \\ \hline
$\psi_i$ & $q^2$ & $\phantom{+}0.15564116060 \cdot 10^{-2}$ & $\psi_i$ & $q^2$ & $\phantom{+}0.15036649108$ \\ \cline{2-3} \cline{5-6}
         & $q^3$ & $-0.48721606804 \cdot 10^{-3}$ & & $q^3$ & $-0.20853110369 \cdot 10^{-1}$ \\ \cline{1-3} \cline{4-6}
$\xi_i$  & $\ell$ & $\phantom{+}0.42329775623 \cdot 10^{-1}$ & $\xi_i$ & $\ell^2/(\ell^2+0.75)$ & $-0.18706065543 \cdot 10^1$ \\ \cline{2-3} \cline{5-6}
         & $\ell^4$ & $-0.28721728138 \cdot 10^{-3}$ & & $\ell^3/(\ell^3+0.75)$ & $-0.71033866423$ \\ \cline{1-3} \cline{4-6}
$\chi_i$ & $q^2 \ell$ & $-0.24519442905 \cdot 10^{-2}$ & $\chi_i$ & $\ell^2 q^2/(\ell^2+0.75)$ & $-0.57586765382$ \\ \cline{2-3} \cline{5-6}
         & $q^2 \ell^4$ & $-0.52409821736 \cdot 10^{-2}$ & & $\ell^3 q^2/(\ell^3+0.75)$ & $\phantom{+}0.68025466334$ \\ \cline{2-3} \cline{5-6}
         & $q^3 \ell^4$ & $\phantom{+}0.14877168217 \cdot 10^{-2}$ & & $\ell^2 q^3/(\ell^2+0.75)$ & $\phantom{+}0.13582356655$ \\ \cline{2-3} \cline{5-6}
         & & & & $\ell^3 q^3/(\ell^3+0.75)$ & $-0.19160979995$ \\ \cline{1-3} \cline{4-6}

\multicolumn{3}{|l|}{$c_2\left(q,\ell \right)$} & \multicolumn{3}{|l|}{$c_3\left(q,\ell \right)$} \\ \hline
$c$      & 1 & $-0.26605902479 \cdot 10^{2}$ & $c$ & & \\ \hline
$\psi_i$ & $q^2$ & $-0.11443846237 \cdot 10^{1}$ & $\psi_i$ & $q^2$ & $-0.10592478758$ \\ \cline{2-3} \cline{5-6}
         & $q^3$ & $\phantom{+}0.10977797047$ & & $q^5$ & $-0.35597308644 \cdot 10^{-2}$ \\ \cline{1-3} \cline{4-6}
$\xi_i$  & $\ell^2/(\ell+0.75)^2$ & $\phantom{+}0.11387292692 \cdot 10^{3}$ & $\xi_i$ & $\ell^{0.5}$ & $-0.88369353198$ \\ \cline{2-3} \cline{5-6}
         & $\ell^3/(\ell+0.75)^3$ & $-0.10829394551 \cdot 10^{3}$ & & & \\ \cline{1-3} \cline{4-6}
$\chi_i$ & $\ell^2 q^2/(\ell+0.75)^2$ & $\phantom{+}0.11316643469 \cdot 10^{2}$ & $\chi_i$ & &  \\ \cline{2-3} \cline{5-6}
         & $\ell^3 q^2/(\ell+0.75)^3$ & $-0.17323579858 \cdot 10^{2}$ & & &  \\ \cline{2-3} \cline{5-6}
         & $\ell^2 q^3/(\ell+0.75)^2$ & $-0.16093695721 \cdot 10^{1}$ & & &  \\ \cline{2-3} \cline{5-6}
         & $\ell^3 q^3/(\ell+0.75)^3$ & $\phantom{+}0.30266683521 \cdot 10^{1}$ & & &  \\ \cline{1-3} \cline{4-6}
\end{tabular}
\end{table}

\clearpage

\listoffigures
\clearpage

\clearpage
\clearpage
\clearpage

\end{document}